\documentclass[twocolumn]{aastex7}

\usepackage{subcaption}
\usepackage{graphicx}
\usepackage{amsmath}
\usepackage{bm}

\begin{document}

\title{Synthetic Fe~\textsc{XIII}~1074.7~nm Observations of Torsional Alfv\'{e}n Waves in Coronal Waveguides}

\author[orcid=0000-0002-3814-4232,gname=Samuel J.,sname=Skirvin]{Samuel J. Skirvin}
\affiliation{School of Engineering, Physics and Mathematics, Northumbria University, Newcastle Upon Tyne, UK}
\email[show]{samuel.j.skirvin@northumbria.ac.uk}  

\author[orcid=0000-0001-5678-9002,gname=Richard J., sname=Morton]{Richard J. Morton}
\affiliation{School of Engineering, Physics and Mathematics, Northumbria University, Newcastle Upon Tyne, UK}
\email{richard.morton@northumbria.ac.uk}  

\author[orcid=0000-0002-7451-9804,gname=Thomas, sname=Schad]{Thomas A. Schad}
\affiliation{National Solar Observatory, Makawao, Hawaii, 96768, USA}
\email{tschad@nso.edu}

\begin{abstract}

Torsional Alfv\'{e}n waves are a promising mechanism for transporting energy through the solar atmosphere, with implications for coronal heating and solar wind acceleration. Recently, signatures of torsional Alfv\'{e}n waves have been observed with the Daniel K. Inouye Solar Telescope. We aim to investigate the effects of line of sight integration and plasma conditions on the observable properties of the torsional mode. Here we present three-dimensional magnetohydrodynamic simulations of multiple coronal waveguides, driven by a combination of transverse kink and torsional wave drivers, considering two plasma regimes representative of active region (AR) and quiet Sun (QS) conditions. Synthetic observables of the Fe~\textsc{xiii}~1074.7~nm coronal emission line are produced using the pyCELP forward-modelling framework to enable direct comparison with spectroscopic observations. In the QS regime, red-blue Doppler asymmetries are associated with the driven torsional waves, though their observed amplitudes are substantially reduced by line-of-sight integration. In contrast, the AR regime exhibits red-blue Doppler asymmetries even in the absence of an imposed torsional driver, which may be misidentified as the $m=0$ torsional Alfv\'{e}n mode. In the AR setup, the red-blue Doppler asymmetries arise from strong phase mixing, generating shear flows and localised vorticity between waveguides where emission is strongest. The differences between the two regimes are governed by the location of peak emission, which is determined by the degree of density inhomogeneity. Our results support the identification of torsional Alfv\'{e}n wave signatures in the quiet Sun (a weakly inhomogeneous environment), but caution should be exercised when interpreting spectroscopic observations in strongly inhomogeneous environments such as active regions.

\end{abstract}

\keywords{Alfv\'{e}n waves (23); Magnetohydrodynamics (1964); Solar atmosphere (1477); Solar corona (1483); Solar coronal waves (1995)}

\section{Introduction}

The heating of plasma in the solar corona directly drives its radiative output in Extreme Ultraviolet (EUV) and X-ray bands \citep{Walsh2003, Klimchuk2006}. Closely related to the issue of coronal heating is understanding the acceleration of solar and stellar winds to hundreds of kilometres per second, a common phenomena of stars with convective envelopes, with broad implications for the habitability of surrounding planets \citep{See2014}.

Magnetohydrodynamic (MHD) waves are leading contenders for explaining plasma heating and solar wind acceleration \citep{arr2015, vanDoorsselaere2020SSRv, morton2023}, with observational ubiquity and modelling suggesting they can deliver enough power to heat some regions of the solar atmosphere. However, previous observations have yet to place significant constraints on the specific physical processes involved, particularly in terms of accurately determining the wave energy flux and the dominant energy damping/dissipation mechanisms \citep[e.g.][]{Hahn2012}. 

There have been great advancements in probing MHD waves in the corona over recent decades with the introduction of coronal spectropolarimeters such as the Coronal Multi-channel Polarimeter \citep[CoMP -][]{Tomczyk2008} and, more recently, Daniel K. Inouye Solar Telescope \citep[DKIST -][]{Rimmele2020} Cryogenic Near-Infrared Spectropolarimeter \citep[CryoNIRSP -][]{Fehlmann2023} which enables detailed analysis of both plane-of-sky (POS) and line-of-sight (LOS) plasma motions at unprecedented resolution. These instruments have revealed prevalent transverse MHD waves in the corona, which can carry energy on the order of $100$~W~m$^{-2}$ \citep[e.g.,][]{Morton2026}, sufficient to accelerate the solar wind and heat the quiet Sun corona. Early reports \citep{Tomczyk_et_al_2007, dep2007, Oka2007} described these waves as Alfv\'{e}n waves (which classically are purely incompressible perturbations of the plasma in a uniform medium of infinite extent), however, follow-up studies \citep{erd2007, van_etal08b} demonstrated that these waves are better described as kink (or Alfv\'{e}nic) waves, which are transverse, largely incompressible waves and whose wavevector is directly parallel to the magnetic field. In an inhomogeneous plasma, such as found in the solar corona, pure (linear) Alfv\'{e}n waves exist solely as torsional motions on individual magnetic surfaces, hence they are not observable through imaging instruments. However, there have only been isolated reports of large-scale Alfv\'{e}n waves in the solar atmosphere using spectrographs, where the waves have been identified through both periodic non-thermal broadening \citep{Jess2009} and anti-phase Doppler velocities \citep{ Kohutova2020}.

Recently, there have been observational reports of small-scale torsional Alfv\'{e}n waves present in coronal structures alongside the commonly observed kink motions. Using  the Fe~\textsc{xiii} $1074.7$~nm emission line observed with DKIST/Cryo-NIRSP, \citet{Morton2026} discovered signatures of small-scale torsional motions in both open field and a quiescent loop, evidenced by the anti-phase Doppler velocity at opposite edges of localised, fine-scale brightness enhancements which appear to highlight the magnetic field (thought to represent over-dense waveguides). The authors extracted these small-scale torsional signatures by subtracting the contribution for the bulk Alfv\'{e}nic (kink) motions at the centre of the identified structures. The torsional Alfv\'{e}n waves were found to have velocity amplitudes on the order of $0.25$~km~s$^{-1}$. However, the true amplitude may be larger, with line-of-sight (LOS) integration of emission from various structures oscillating out-of-phase leading to substantially smaller Doppler velocities \citep[see, e.g., ][for the case of kink modes]{Pant2019}. The study by \citet{Morton2026} inferred, through Monte Carlo simulations based on linear wave theory, that the true velocity amplitude of the small-scale Alfv\'{e}n waves may be around $20$~km~s$^{-1}$. The spatial scales of the torsional Alfv\'{e}n waves were found to be on the order of $1-2$~Mm, confined to individual brightness enhancements. While some isolated events displayed torsional signatures extending beyond $5$~Mm, this is in contrast to measurements of kink modes that are coherent over scales of $\sim$14~Mm \citep{Sharma2023} measured by CoMP. 

\medskip 

The aim of this paper is to numerically model the presence of both kink and small-scale torsional Alfv\'{e}n waves in coronal waveguides and explore two distinct plasma regimes to understand the spectroscopic observability of torsional waves in different plasma environments in the solar atmosphere. Moreover, we also discuss the effect of LOS integration on the Doppler velocity amplitudes associated with these waves.  This paper is structured as follows: in Section \ref{sec:methods} we outline the setup of the numerical model and wave drivers, in addition to presenting the forward modelling techniques utilised. In Section \ref{sec:results} we present results from the simulation and compare the forward modelled output from both setups and identify the origins of red/blue asymmetries. Finally, in Section \ref{sec:conclusions} we summarise our findings and discuss avenues for follow-up studies.

\section{Methods}\label{sec:methods}
\subsection{Numerical Setup}\label{subsec:setup}

To model the wave dynamics in coronal waveguides we conduct a 3D MHD simulation using the PLUTO code \citep{Mign2007, Mign2012, Mign2018}. We solve the ideal MHD equations and use the Harten-Lax-Van Leer (HLL) approximate Riemann solver, with a piece-wise total variation diminishing (TVD) linear reconstruction method for the spatial integration. For time advancement, a second order Runge-Kutta scheme is used with Strang operator splitting \citep{Strang1968}. The solenoidal constraint is controlled by the eight wave formulation described by \citet{Powell1999}. Thermal conduction along the magnetic field is included and applied using the super-time-stepping technique. Radiative losses and external heating rates are not taken into account, as the associated cooling timescales of $\sim 20$~minutes \citep[e.g.][]{Aschwanden2008} are comparable to the simulation duration, and significant thermal evolution of the background plasma is not expected. For simplicity, we also assume full plasma ionisation throughout the whole domain. Unlike the study by \citet{Pant2019}, we ignore gravity in the current model for simplicity, which results in the coronal waveguides having a density which is invariant with height.

\begin{figure*}
    \centering
    \includegraphics[width=0.99\linewidth]{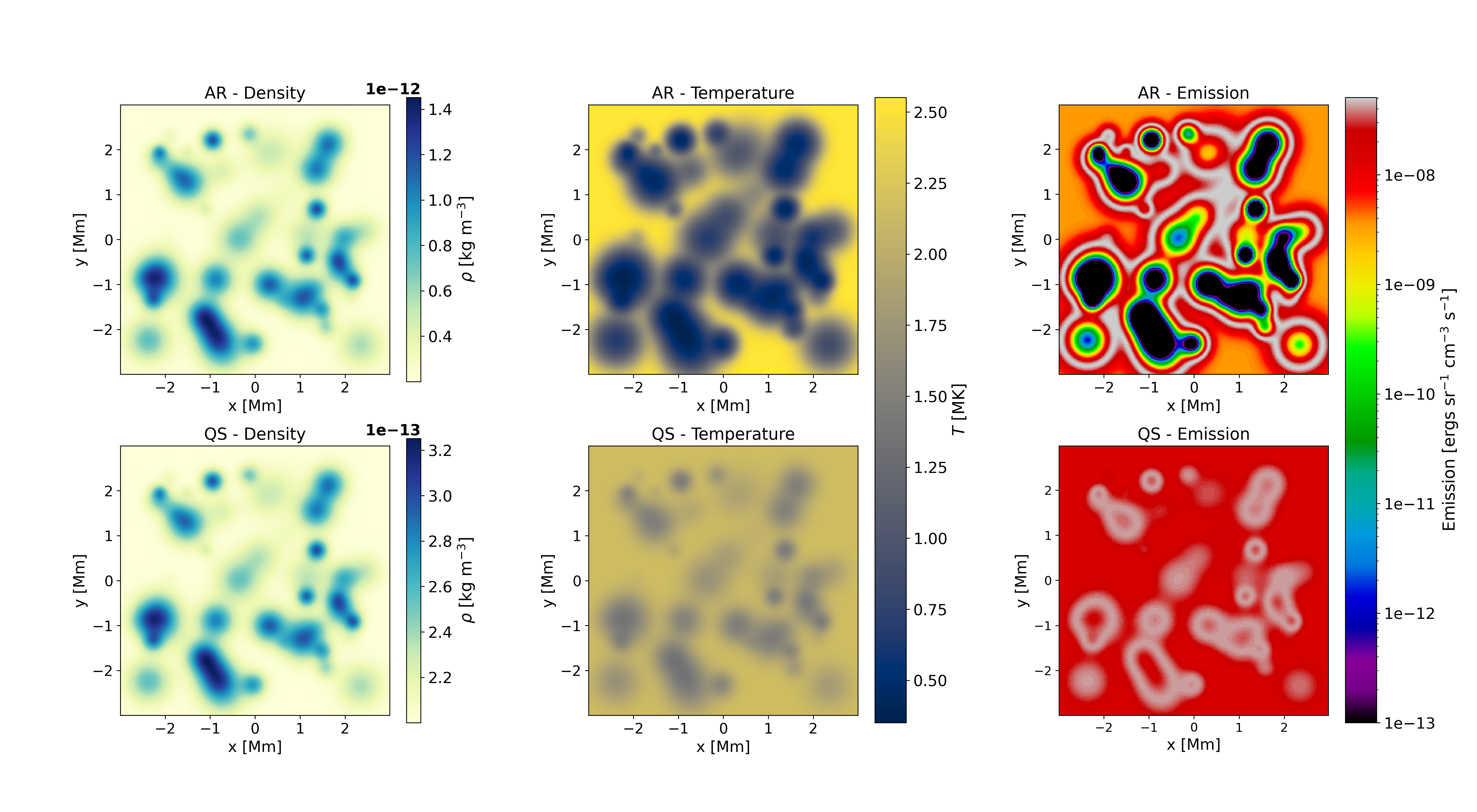}
    \caption{Horizontal slices at the bottom boundary of the simulation for the AR simulation setup (top row) and the QS simulation setup (bottom row) - see discussion in Section \ref{subsec:setup}. The left panels show the plasma density whereas the temperature is displayed in the middle panels. The right hand panels show the line integrated emission of the Fe~\textsc{xiii} 1074.7~nm spectral line (see discussion in Section \ref{subsec:emission}).}
    \label{fig:initial_setup}
\end{figure*}

The simulations are conducted in a Cartesian geometry ($x,y,z$), on a uniform grid without any refinement. The grid size is $512 \times 512 \times 128$ that spans $6 \times 6 \times 50$ Mm, resulting in a grid resolution of $11.7 \times 11.7 \times 390.6$~km. We create overdense waveguides by placing $50$ Gaussian enhancements in plasma density at random locations in the $x$-$y$ plane, which creates transverse inhomogeneities to guide the excited MHD waves. The plasma density is prescribed according to the following equation:
\begin{equation}
    \rho(x,y) = \rho_0 + \sum_{i=0}^{50} A_i \exp^{-\frac{\left((x-x_i)^2+(y-y_i)^2\right)}{2\sigma_i^2}},
\end{equation}
where $\rho_0 = 2\times 10^{-13}$ kg~m$^{-3}$ (which is equivalent to an electron density $n_e \approx 1\times 10^{8}$~cm$^{-3}$ assuming a Hydrogen:Helium abundance ratio of 10:1). We consider two different simulation setups, varying the amplitude of the overdense waveguides to change the density ratio (ratio of density at the waveguide centre to the ambient plasma). We shall denote the two simulation setups as (1) ``AR'' for the strong density contrast simulation (representative of an active region) and (2) ``QS'' for the smaller density ratio (quiet Sun). The magnitude of the inhomogeneity is given by $A_i$ which is chosen from a uniform distribution of $[0,5] \ \rho_0$ for the AR simulation and a uniform distribution of $[0,0.5] \ \rho_0$ for the QS simulation. These values are chosen based upon estimates for the density contrast from observations \citep{Arregui2013, Arregui2014, Pascoe2017, Morton2021_QS_damping}. Ultimately, the background plasma density and temperature for both the AR and QS setups possess values of $2\times 10^{-13}$~kg~m$^{-3}$ and $2-2.5$~MK, respectively, and is in agreement with determinations using direct and white-light imaging \citep{Esser1999, Milligan2005}, radio measurements \citep{Mercier2015} and DEM analysis \citep{Warren2003}. The maximum density (minimum temperature) in the QS setup reaches $3.2\times 10^{-13}$~kg~m$^{-3}$ and $\approx 1$~MK, respectively, which matches high resolution EUV observations of the quiet Sun \citep{Winebarger2013, Dolliou2023}. On the other hand for the AR setup, the maximum density (minimum temperature) reaches $1.4\times 10^{-12}$~kg~m$^{-3}$ ($n_e\approx7.2\times 10^{8}$~cm$^{-3}$) and $\approx 0.5$~MK, respectively, consistent with observations of the multi-thermal nature of `warm' active region loops \citep{DelZanna2011}.

The spatial extent of the inhomogeneity is kept consistent across both simulations and is controlled by $\sigma_i$ which is randomly selected from $[100, 350]$~km and produces circular enhancements. To avoid numerical issues with overdense enhancements at the boundaries, coupled with periodic boundary conditions, the Gaussian enhancements are placed within a sub domain ranging from $[-2.5,2.5]$~Mm and selected randomly from a uniform distribution. This ensures that the Gaussian enhancements do not overlap the boundary which would then require numerical relaxation to reach an equilibrium state. A snapshot of the initial domain for both simulation setups is displayed in Figure \ref{fig:initial_setup}.

To ensure the numerical setup is in pressure equilibrium, the density enhancements are balanced by a decrease in plasma temperature to maintain thermal pressure balance. To maintain reasonable temperatures motivated by coronal observations, the prescribed gas pressure varies across the two simulations. The magnetic field is taken to be straight and uniform everywhere, directed along $z$ with a magnitude of $5$~G. This value of the magnetic field is typical for the quiet Sun \citep{Yang2024Sci}, and results in an ambient plasma-$\beta$ of $0.04$. It should be noted that the magnetic field strength is kept constant at $5$~G for both simulation setups. Fixing the magnetic field strength allows us to isolate the effects of density structuring and thermodynamic differences on wave propagation and observability, without introducing additional scaling from magnetic amplitude.

The boundary conditions are set to be periodic in $x$ and $y$, however, are left open at the upper boundary to allow waves to leave the domain. At the bottom boundary, the $x$ and $y$ components of the velocity are set as described in Section \ref{sec:driver}, whereas all other variables use a zero-gradient (outflow-like) extrapolation from the first active zone into the ghost cells.

\begin{figure}
    \centering
    \includegraphics[width=0.9\linewidth]{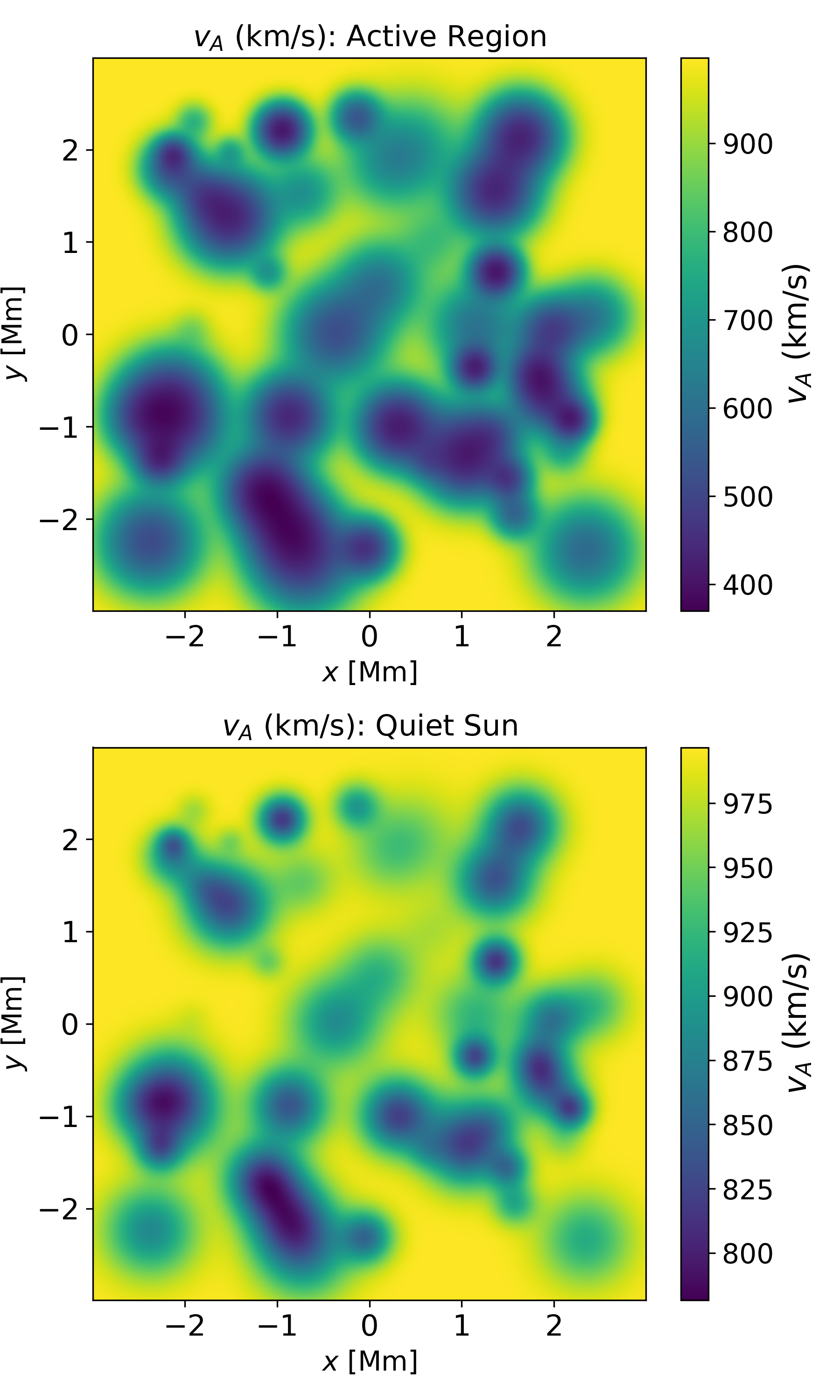}
    \caption{Cross section of the Alfv\'{e}n speed, $v_A$ in the AR simulation (top) and QS simulation (bottom). As gravity is neglected in the current work, the Alfv\'{e}n speed is uniform with height.}
    \label{fig:valf_comparison}
\end{figure}

Figure \ref{fig:valf_comparison} displays a cross-section at the bottom boundary of the Alfv\'{e}n speed for both simulation setups. It is evident that the magnitude of the Alfv\'{e}n speed varies between the two setups, with stronger transverse gradients present in the AR simulation.

\subsection{Details of wave drivers}\label{sec:driver}

\begin{figure}
    \centering
    \includegraphics[width=0.99\linewidth]{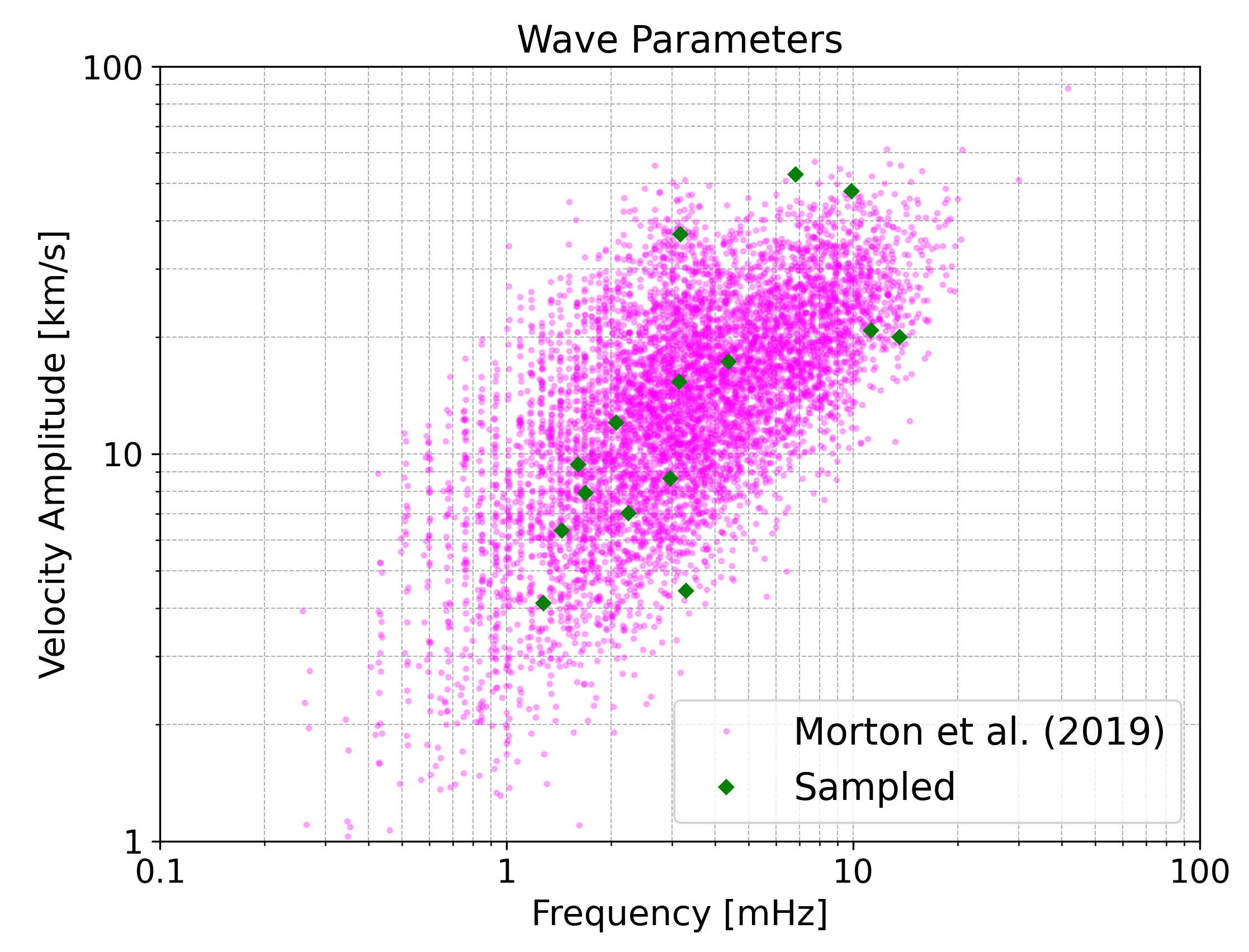}
    \caption{Distribution of the observed velocity amplitude and wave frequencies from \citet{Morton2019} shown in magenta and the selected parameters used for the transverse waves in this study displayed in green.}
    \label{fig:wave_params}
\end{figure}

To disentangle the combined effects from the driven transverse and torsional waves, we perform two separate simulations. One simulation contains both the torsional drivers and transverse drivers (`torsional with transverse'), whereas the second simulation has just the transverse wave drivers (`transverse only'). The latter is similar to the study by \citet{Pant2019}.

\medskip

We excite transverse waves by perturbing the entire bottom boundary through introducing motions in the $x$ and $y$ velocity components (i.e., perpendicular to the direction of the background magnetic field) through a superposition of 15 different velocity drivers with different amplitude, periodicity and phase. The velocity amplitudes and wave frequencies are sampled from a measured distribution \citep{Morton2019} and displayed in Figure \ref{fig:wave_params}. The $x$ and $y$ components of the velocity driver are given by the following equations:
\begin{equation}\label{eqn:vx_trans}
    v_{x, \text{trans}}(z=0, t) = \sum_{i=1}^{15} U_i \sin (\omega_i t + \phi_i),
\end{equation}
\begin{equation}\label{eqn:vy_trans}
    v_{y, \text{trans}}(z=0, t) = \sum_{i=1}^{15} V_i \sin (\omega_i t + \phi_i),
\end{equation}
where $\phi_i$ is a random phase between $[0,2\pi]$ selected from a uniform distribution. Unlike \citet{Pant2019}, the periods (and hence the $\omega$'s) and the velocity amplitudes are chosen from an observed power law distribution \citep{Morton2019} using a bivariate log-normal model to ensure that the sampled parameters are statistically consistent with the observed frequency-amplitude distribution. Figure \ref{fig:wave_params} displays the observed relationship between velocity amplitude and frequency for the Alfv\'{e}nic waves in addition to the selected values considered in this work, displayed in green. Once the waves are superimposed, the RMS of the resulting transverse wave signal is around $3$~km~s$^{-1}$ which is representative of observed values from POS motions \citep{Thurgood2014, Morton2026}. The spatial scale of the driver is consistent with observations that indicate kink motions are coherent over patches of around $5-14$~Mm perpendicular to the field \citep{Sharma2023, Hahn2025}.

\medskip

We also excite torsional waves at the bottom boundary. The plasma in the intergranular lanes regularly displays vortical motions, which are expected to launch torsional waves/pulses \citep{Battaglia2021, Tziotziou2023SSRv, Kannan2024}. The torsional waves are introduced at (seven) specific locations which are chosen to lie within density enhancements, centred on $r_{0,i}$ for each driver. In observations, the density enhancements outline the magnetic field in the corona, which is ultimately rooted in the photosphere where torsional motions are present. The torsional wave drivers are prescribed at the bottom boundary $z=0$ according to the following expression:

\begin{equation}
    f = \sqrt{(x-x_{c})^2 +(y-y_{c})^2},
\end{equation}
\begin{align}\label{eqn:vx_tors}
      v_{x, \text{tors}}(t) &= \sum_{i=1}^{7} \frac{-W_i (y-y_{c,i})}{f_i}  \exp\left(\frac{-(f_i-r_{0,i})^2}{2 \sigma^2}\right) \times \\ \nonumber & \times
    \sin(\omega_{\text{tors}}t + \phi_{i, \text{tors}}),  
\end{align}
\begin{align}\label{eqn:vy_tors}
      v_{y, \text{tors}}(t) &= \sum_{i=1}^{7} \frac{W_i (x-x_{c,i})}{f_i}  \exp\left(\frac{-(f_i-r_{0,i})^2}{2 \sigma^2}\right) \times \\ \nonumber & \times
    \sin(\omega_{\text{tors}}t + \phi_{i, \text{tors}}),  
\end{align}
where $\sigma=25$~km controls the width (radius) of the driver and $\omega_{\text{tors}}$ is the frequency of the torsional drivers, which is taken to be $P=2\pi/\omega_{\text{tors}} = 150$~s. The amplitude of the torsional waves is chosen to be significantly larger than the transverse motions and we set $W_i=20$~km s$^{-1}$ which is scaled randomly by $82\% - 114\%$ to create drivers with varying amplitude around the observationally inferred value \citep{Morton2026}. This results in torsional drivers with velocity amplitude ranging from $W_i = 16.4 - 22.8$ km s$^{-1}$. The centres of the torsional drivers are taken as $x_{c,i}$ and $y_{c,i}$ and are picked to match the centre of specific overdense waveguides chosen by eye. Each torsional driver is also given a random phase shift taken from a uniform distribution between $\phi_{\text{tors}}=[0,2\pi]$. The total perturbation applied at the bottom boundary is then a simple summation of Equations (\ref{eqn:vx_trans}) \& (\ref{eqn:vx_tors})  and Equations (\ref{eqn:vy_trans}) \& (\ref{eqn:vy_tors}), i.e. $v_{\text{trans}} + v_{\text{tors}}$ for each component. The factor $f_i^{-1}$ in Equations~(\ref{eqn:vx_tors}) and~(\ref{eqn:vy_tors}) is formally singular at $f_i=0$; however, this singularity is not encountered in practice since the Gaussian envelope $\exp\left[-(f_i - r_{0,i})^2 / 2\sigma^2\right]$ ensures the driver amplitude is vanishingly small near $f_i = 0$, as the driver is designed to excite torsional motions on an annulus of radius $r_{0,i} \gg 0$ rather than at the centre of each waveguide.

\subsection{Forward Modelling}
To investigate the LOS integration effects on the observable wave properties, we perform forward modelling of the forbidden Fe~\textsc{xiii} $1074.7$~nm infrared line using the "python package for Coronal Emission Line Polarization calculations" (pyCELP) code \citep{Schad2020, Schad2021}. pyCELP solves a set of statistical equilibrium equations in the spherical statistical tensor representation for a multi-level atom for the no-coherence case. This approximation is useful in the case of forbidden line emission by visible and infrared lines, such as Fe~\textsc{xiii} $1074.7$~nm. pyCELP calculates the atomic density matrix elements for a single ion under coronal equilibrium conditions and excited by a prescribed radiation field and thermal collisions. The output from pyCELP consists of polarized Stokes emissivities for each simulation grid cell, which are used to synthesize emergent line profiles along each optically thin line of sight. These profiles are then used to determine effective LOS Doppler velocities and line widths.

\begin{figure}
    \centering
    \includegraphics[width=0.89\linewidth]{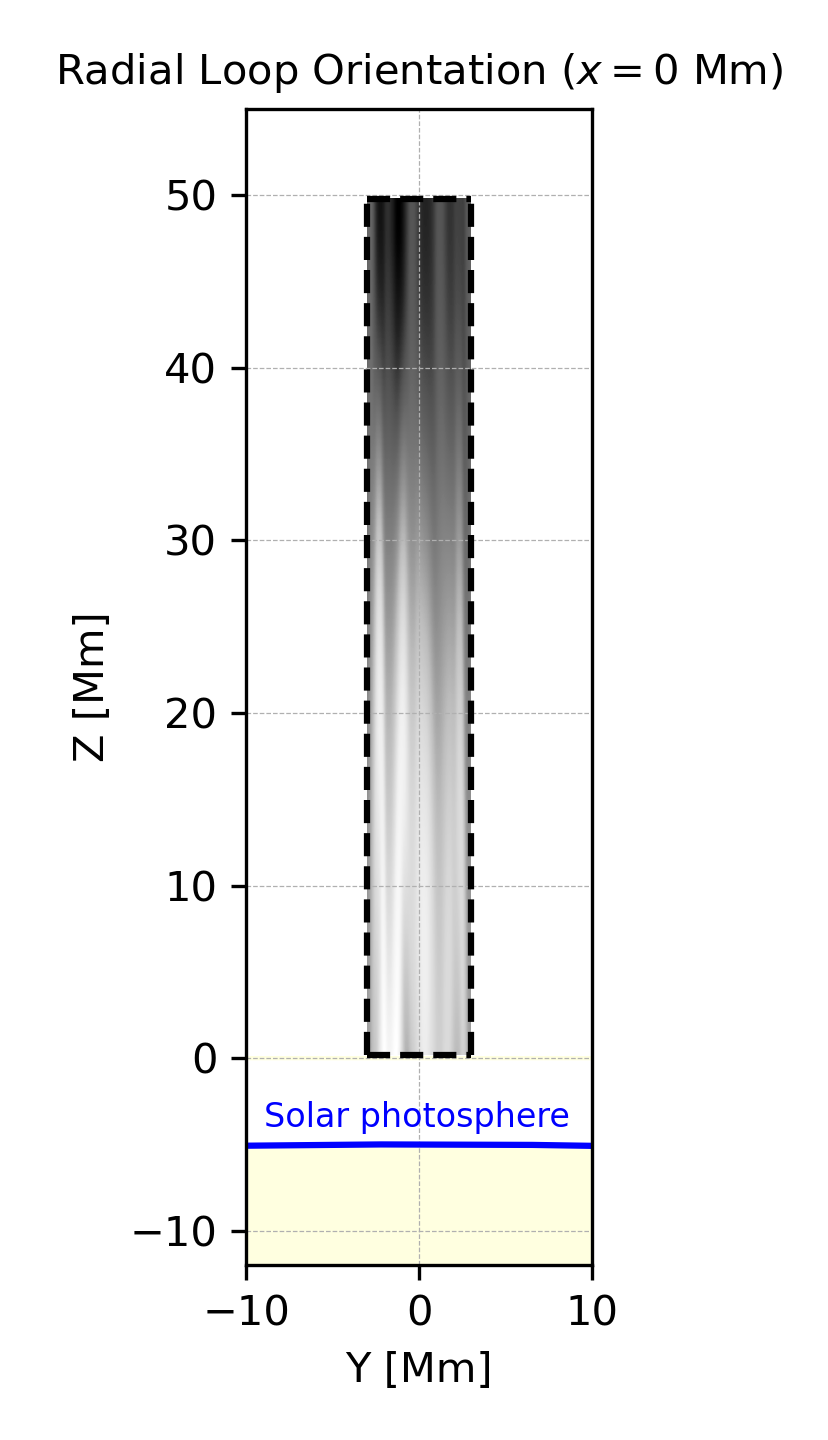}
    \caption{Radial loop orientation used to conduct the forward modelling with pyCELP. The diagram displays a snapshot of the $x$ component of the magnetic field, $B_x$, at $t=300$~s.} 
    \label{fig:radial_orientation}
\end{figure}

For the discussion in this paper, we will assume that the waveguides are perpendicular to the solar surface (i.e., radial) as outlined in Figure \ref{fig:radial_orientation}. The intensity of the Fe~\textsc{xiii} 1074.7~nm line includes a contribution from photo-excitation by the photospheric radiation field, whose strength decreases with height above the solar surface. We therefore specify the height of the loop's bottom boundary to be $z_0=5$~Mm to appropriately account for the degree of radiative excitation at the location of the simulated loop. For simplicity, we don't recompute the ionisation calculations at each height in the simulation domain, as the densities in the model result in collisional excitation dominating over photoexcitation. Therefore, the location of $z=0$ in the numerical simulation, corresponds to forward modelling coordinates $z = -(R_{\odot} +5)$~Mm. Finally, we take the LOS direction to be the $x$-axis in the simulation. For reference, the peak ionization formation temperature of the Fe~\textsc{xiii} 1074.7~nm line is $\sim 10^{6.25}$~K

For the forward modelling, to speed up computation, the spatial resolution is reduced by a factor of 2 in each dimension which results in a horizontal spatial resolution of $22$~km. This resolution is still $4$ times greater than the resolution of Cryo-NIRSP and significantly higher resolution than CoMP.

\medskip 

For a given spectral line, the LOS Doppler shift ($\lambda_D$) is computed from the first moment:

\begin{equation}\label{eqn:Doppler_Shift}
\lambda_D(z, y, t)= 
\frac{\displaystyle \int I(\lambda, z, y, t) \, \lambda \, d\lambda}
     {\displaystyle \int I(\lambda, z, y, t) \, d\lambda} - \lambda_0,
\end{equation}

\noindent
where  $\lambda_0$ is the location of the central peak of the Fe~\textsc{xiii} emission line, which is $1074.7$~nm, \( I(\lambda, z, y, t) \) is the Stokes \( I \) intensity as a function of wavelength, height, position, and time. The Doppler velocity is then computed through converting $\lambda_D$ from units of wavelength to units of velocity. The line-width is given by the central second moment:

\begin{equation}
% --- Line width (velocity dispersion) ---
\sigma_\lambda^2(z, y, t)=
\frac{\displaystyle \int I(\lambda, z, y, t)
\left[ \lambda - \lambda_D(z, y, t) \right]^2 d\lambda}
{\displaystyle \int I(\lambda, z, y, t) \, d\lambda}, \\[1em]
\end{equation}

\noindent
where \( \sigma_\lambda \) is the intensity-weighted standard deviation of the line profile in wavelength space.

\subsection{QS vs AR Emission}\label{subsec:emission}

Since the Fe~\textsc{xiii} emission line is sensitive to both coronal temperature and electron density, the forward-modelled intensity is expected to vary considerably between the two numerical setups. The right hand panels in Figure \ref{fig:initial_setup} display the wavelength integrated emission. The wavelength-integrated emission is governed by the Fe~\textsc{xiii} contribution function, which depends on local temperature and electron density. It is evident that there are significant differences in emission between the two environments. The AR setup contains dense-cool waveguides, with temperatures around $0.5$~MK and density ratios (between internal and external plasma) which peak around $\rho_i/\rho_e = 7$. In the AR setup, the emission from overdense waveguides is around $5$ orders of magnitude weaker in Fe~\textsc{xiii} 1074.7~nm than the surrounding plasma, meaning the observed emission is dominated by the external medium and the waveguides themselves are only a minor contribution to the total emission. In contrast, the QS setup produces strong Fe~XIII emission from the overdense waveguides, such that the synthetic observations directly sample the waveguides. This makes the QS setup considerably more favourable for associating observed wave signatures with the overdense structures that guide them.

While we have used specific values for temperature and density here, the result is more general. In regions of the corona where there are large differences between the ambient and over dense loop plasmas, unsurprisingly the emission will be dominated by the plasma where the Fe~XIII response is strongest. This may only be a small volume of the plasma or a volume between the waveguides. While in a plasma where the magnitude of inhomogeneity is small, like the quiet Sun and open-field regions, the total emission will have meaningful contributions from a larger volume of plasma.

\section{Results}\label{sec:results}

\subsection{Driven Waves}

\begin{figure*}
    \centering
    \includegraphics[width=0.99\linewidth]{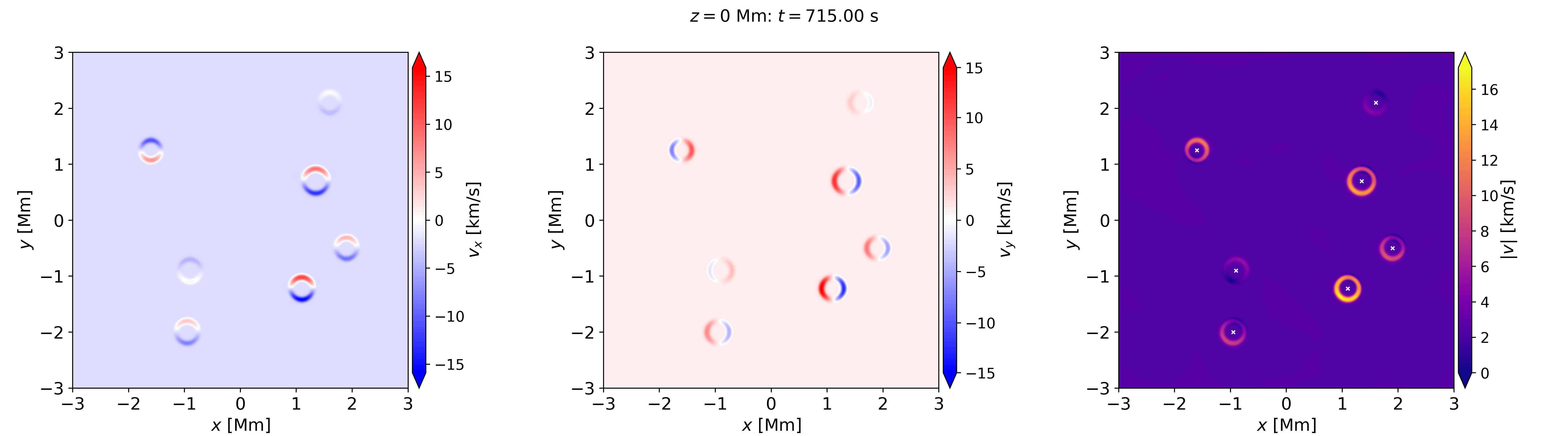}
    \caption{Left: $v_x$, Middle: $v_y$ and Right: $|v|$ (magnitude) at the bottom boundary of the simulation domain at $t=715$~s. The white crosses in the right panel outline the locations of the seven torsional wave drivers in this work.} 
    \label{fig:vels_t715}
\end{figure*}

A snapshot of the velocities at the bottom boundary in the plane transverse to the magnetic field ($x-y$) at $t=715$~s is shown in Figure \ref{fig:vels_t715} for the simulation where the waveguides are perturbed by both the torsional and transverse wave drivers. The locations of the torsional wave drivers are evident through their stronger velocity amplitudes, moreover, Figure \ref{fig:vels_t715} highlights the random phase applied to each individual torsional wave driver. Although the driver does not perturb the $z$ component of the velocity, longitudinal waves are produced through the mode coupling in non-uniform plasmas between transverse and longitudinal waves. The $v_z$ motions are most prominent inside the non-uniform boundary layers of each individual density enhancement. These waves are confirmed as acoustic in nature as they travel at the local sound speed. However, the longitudinal motions do not affect the analysis in this work as they remain well within the linear regime ($v_z \ll c_s$), where $c_s$ is the local sound speed, and the forward modelling does not consider any inclined angles with respect to the magnetic field. \citet{Morton2026} discussed the possibility of opposite field-aligned flows potentially creating torsional Doppler signatures across a coronal loop, and, as stated, we avoid such a possibility by forward modelling perpendicular to the magnetic field ($z$-axis) at all times.

As the torsional drivers are not positioned directly at the centre of mass of the density enhancements, they also excite localised transverse waves, albeit with a very small amplitude compared to the driven bulk transverse waves. This is not unexpected, and the breaking of azimuthal symmetry of magnetic flux tubes has been shown to excite transverse waves previously from compressible wave drivers \citep{Skirvin2023ApJ, Gao2023}. Given the inhomogeneity in the background model, especially in the AR setup, it is expected that phase mixing is present throughout the simulation due to the strong gradient in the Alfv\'{e}n speed (e.g. see Figure \ref{fig:valf_comparison}). 

\subsection{Fe~\textsc{xiii} Forward Modelling}\label{sec:results_pycelp}

\begin{figure*}   
\centering
\begin{subfigure}{0.7\textwidth}
    \includegraphics[width=\textwidth]{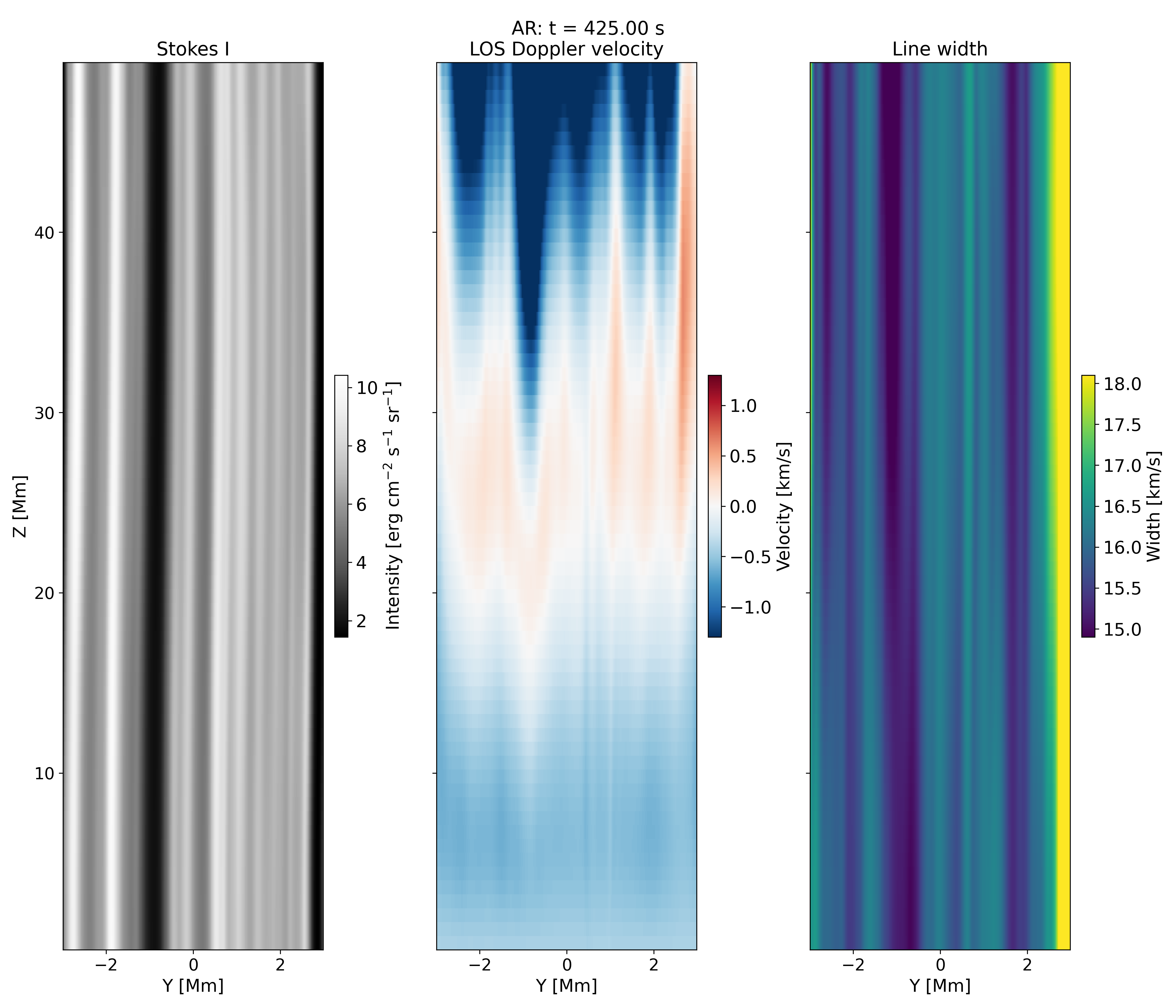}
    \caption{ }    
    \label{fig:pycelp_AR_snapshot}
\end{subfigure}
\begin{subfigure}{0.7\textwidth}
    \includegraphics[width=\textwidth]{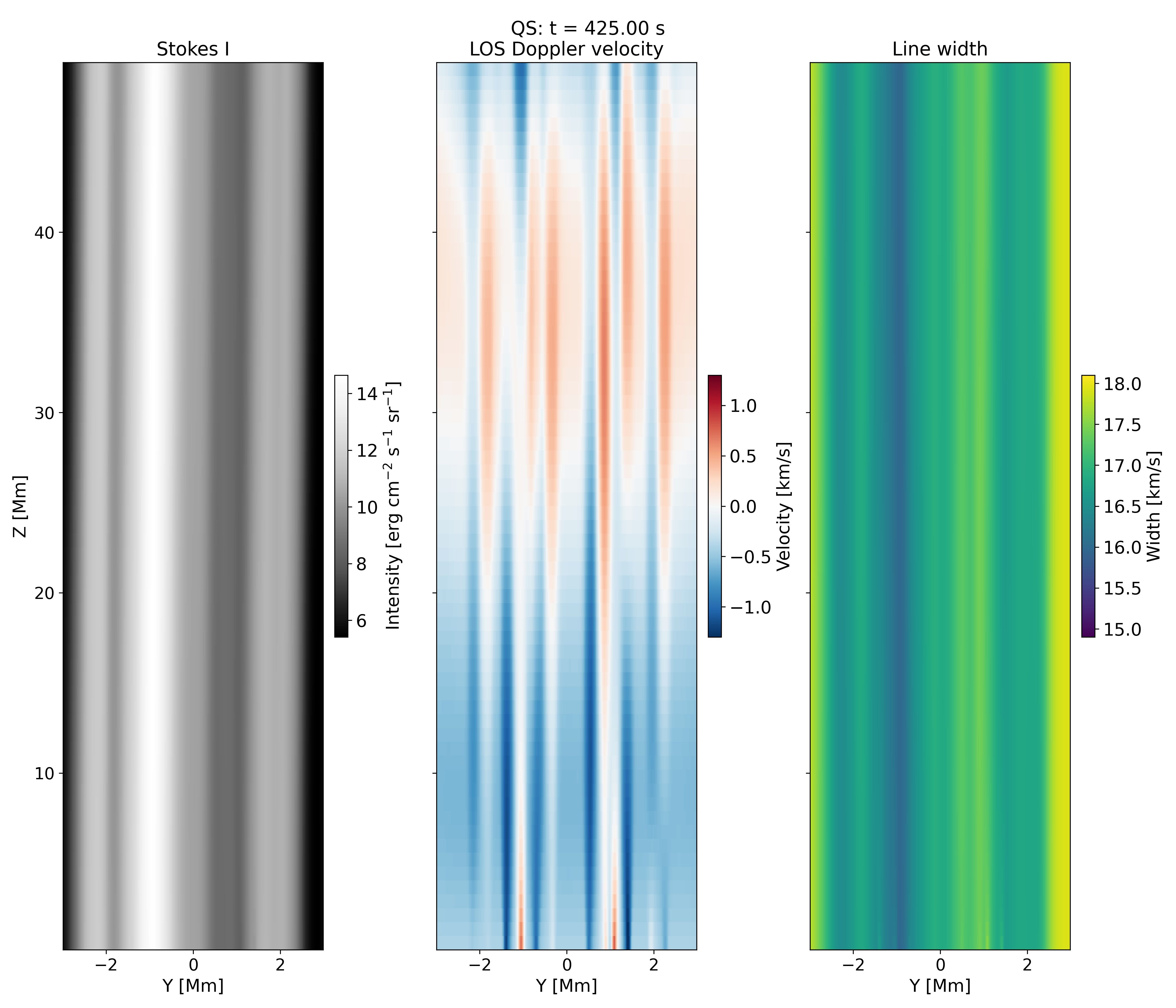}
    \caption{}
    \label{fig:pycelp_QS_snapshot}
\end{subfigure}
\caption{A snapshot at $t=425$~s for the synthesised forward model of the Fe~\textsc{xiii} $1074.7$~nm emission line for a LOS angle of $0^{\circ}$ for (a) the AR simulation and (b) the QS simulation. Left: LOS integrated Stokes intensity. Middle: LOS Doppler velocity and right: total line-width.} \label{fig:pycelp_snapshot}
\end{figure*}

A snapshot of the LOS integrated Stokes intensity, LOS Doppler velocity and Fe~\textsc{xiii} $1074.7$~nm line-width at $t=425$~s in both the AR and QS simulations, for a LOS angle corresponding to $0^{\circ}$ with respect to the observers LOS, are displayed in Figure \ref{fig:pycelp_snapshot}. It is evident how the integrated emission along the LOS (taken to be the $x$-axis in the simulation) appears different in both the AR and QS simulation setups, as expected from Figure \ref{fig:initial_setup}. As a result, the Doppler and line width also appear largely different in both simulations. Regions which appear dark in the AR simulation appear bright in the QS simulation, and vice-versa. Nonetheless, the amplitudes of the Doppler velocity and values for the line-width are consistent across both simulations. It is also worth noting that the obtained line widths are typically larger for the QS case compared to the AR case, which is consistent with observations \citep{Chae1998, Brooks2016}.

\begin{figure*}
\centering
\begin{subfigure}{0.49\textwidth}
    \includegraphics[width=\textwidth]{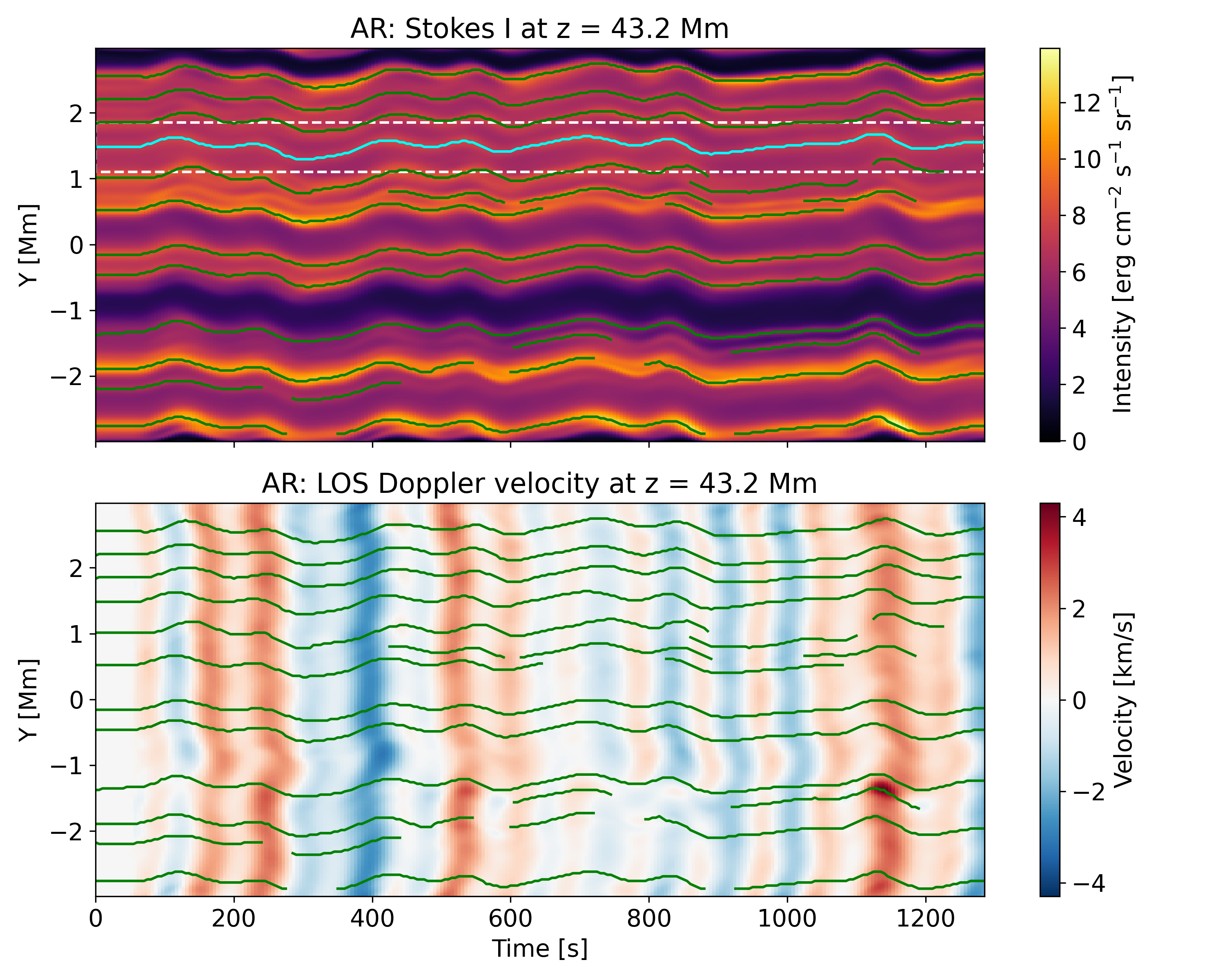}
    \caption{AR simulation}
    \label{fig:NUWT_td_full_AR}
\end{subfigure}
\begin{subfigure}{0.49\textwidth}
    \includegraphics[width=\textwidth]{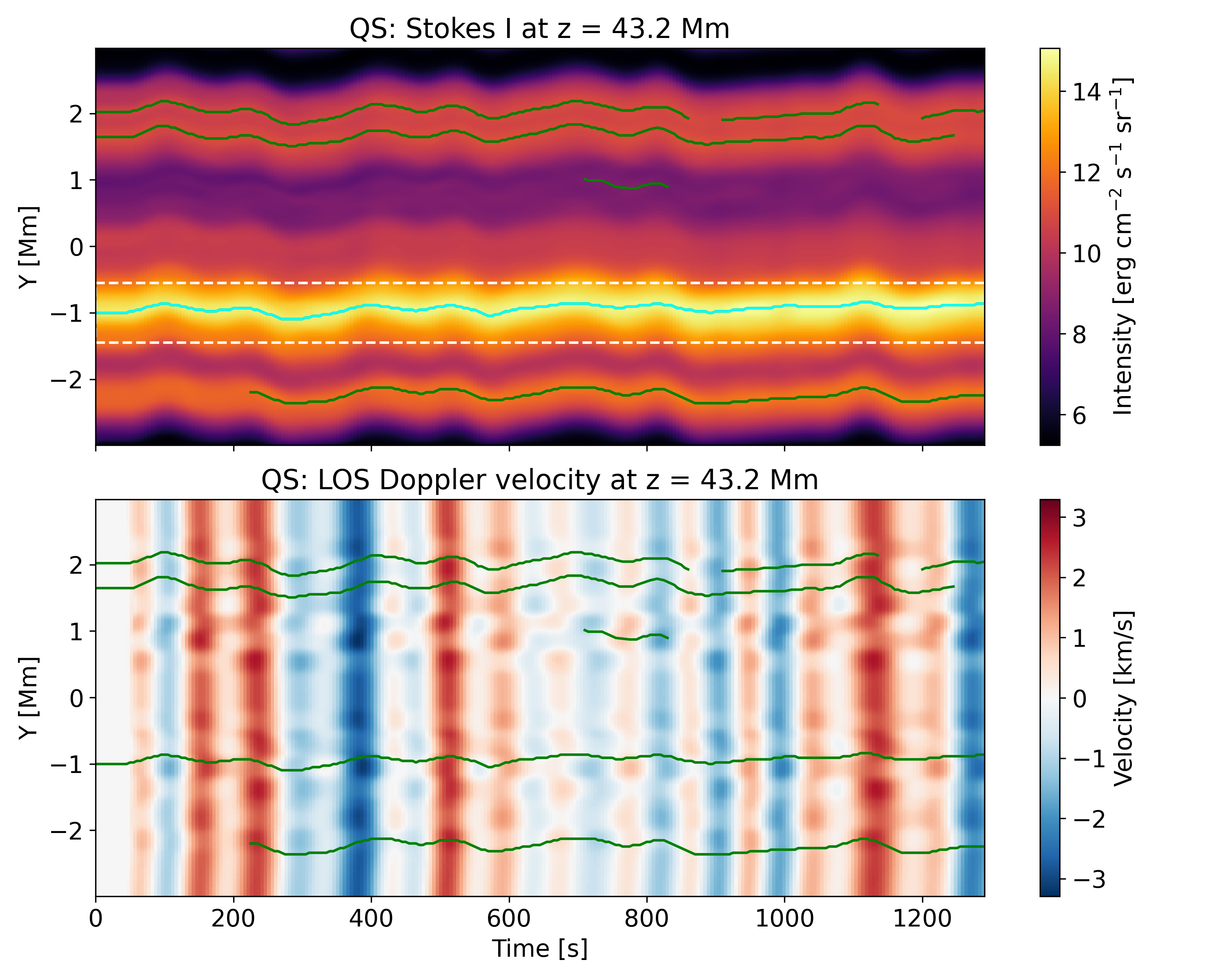}
    \caption{QS simulation}
    \label{fig:NUWT_td_full_QS}
\end{subfigure}
\caption{Time distance diagrams at a height of $z=43.2$~Mm of Stokes I and LOS Doppler velocity for the Fe~\textsc{xiii} $1074.7$~nm line. Figure~\ref{fig:NUWT_td_full_AR} shows the full time-distance plots for the AR simulation. There green curves highlight the threads identified by the NUWT algorithm where the dashed white lines (and solid cyan line) outline the sub-domain for the zoom-in plot in Figure~\ref{fig:NUWT_td_zoom_AR} used for further analysis. Figure~\ref{fig:NUWT_td_full_QS} displays the same but for the QS simulation.}
\label{fig:NUWT_tds}
\end{figure*}

To extract the transverse and torsional motions from the forward modelled output, we mimic sit-and-stare data by placing a horizontal (transverse to the magnetic field) slit across the POS to construct a time-distance diagram at a height of $z=43$~Mm for the Stokes I, LOS Doppler velocity and line-width. The result of this is displayed in Figure \ref{fig:NUWT_tds}. 

\begin{figure*}
\centering
\begin{subfigure}{0.49\textwidth}
    \includegraphics[width=\textwidth]{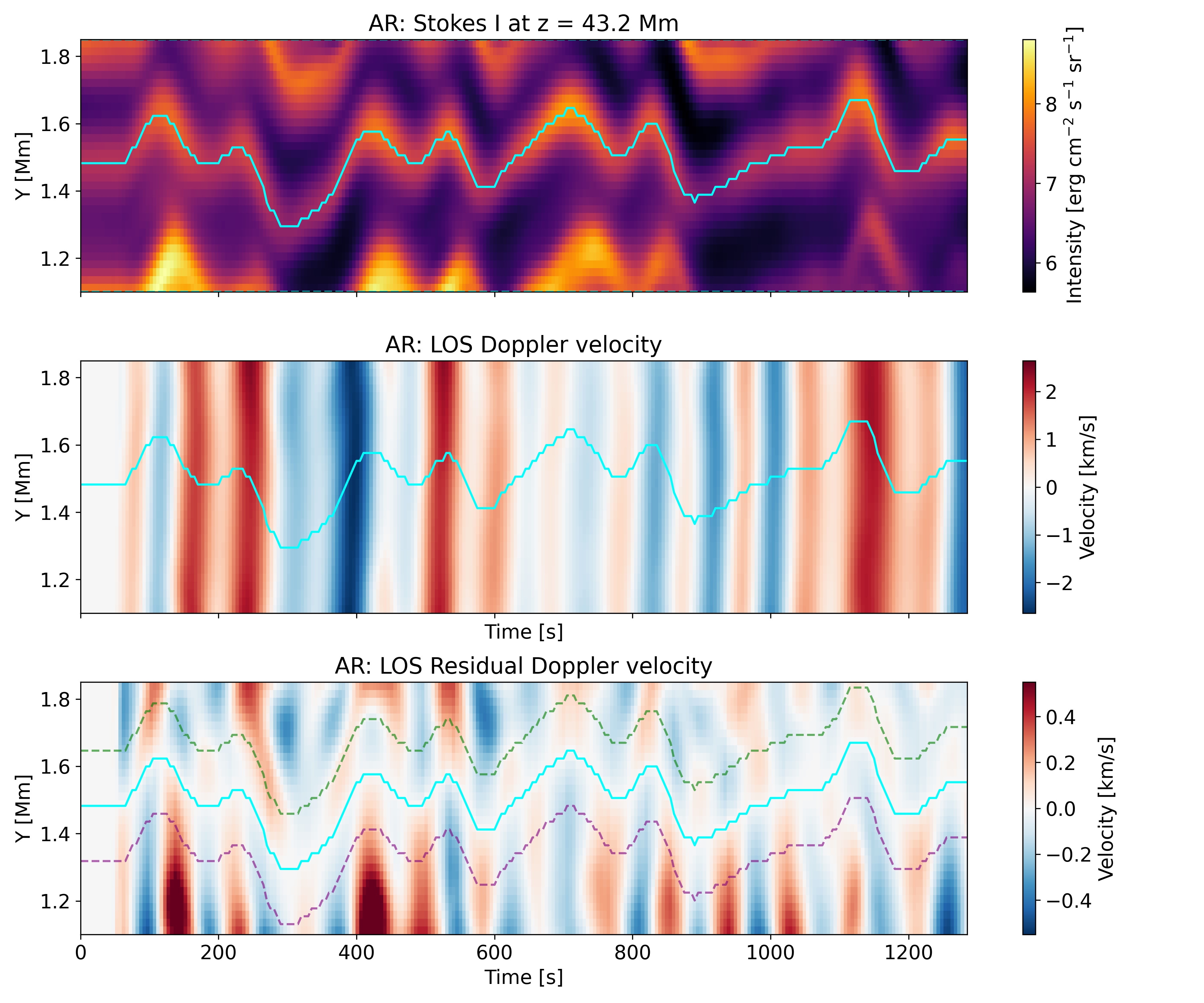}
    \caption{AR simulation}    
    \label{fig:NUWT_td_zoom_AR}
\end{subfigure}
\begin{subfigure}{0.49\textwidth}
    \includegraphics[width=\textwidth]{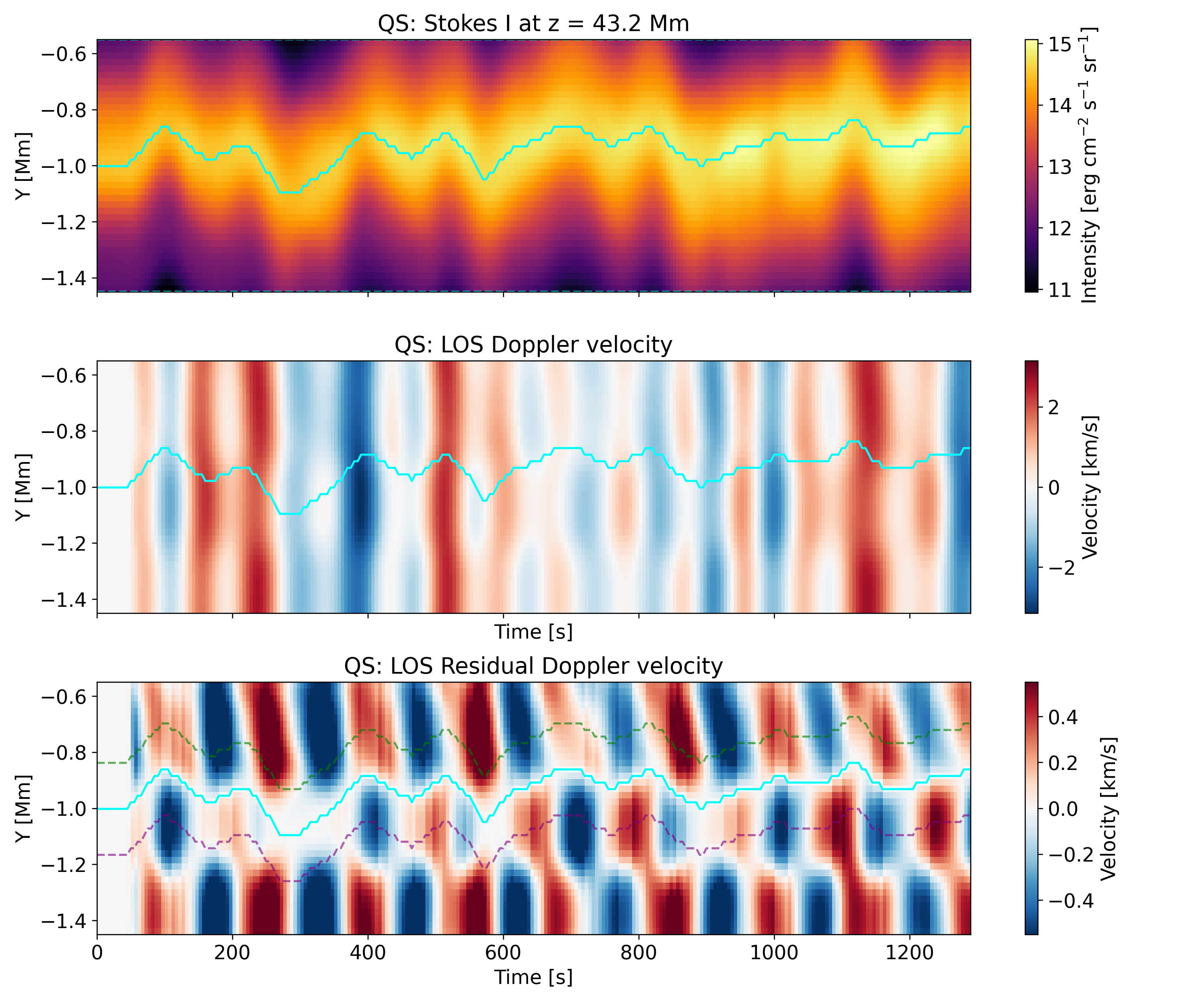}
    \caption{QS simulation}
    \label{fig:NUWT_td_zoom_QS}
\end{subfigure}
\caption{Time distance plots for the zoomed-in region highlighted by the dashed white lines in Figure \ref{fig:NUWT_tds} for (a) the AR simulation and (b) the QS simulation. The top panels show the Stokes Intensity, the middle panels show the full Doppler velocity, whereas the bottom panels highlight the residual Doppler velocity, obtained by subtracting the Doppler velocity along the cyan line above. The dashed purple and green curves in the bottom panels indicate the locations $\pm 7$ pixels from the centre of the chosen thread, corresponding to roughly $150$~km from the centre of the thread.}
\label{fig:NUWT_zoom_tds}
\end{figure*}

Figure \ref{fig:NUWT_tds} shows the resulting time-distance diagram for both the AR and QS simulations with both torsional and transverse wave drivers superimposed. The top panels display the Stokes intensity across the horizontal slit, and the green lines indicate the waveguides (`threads') identified by the NUWT algorithm \citep{Weberg2018}. The NUWT code identifies regions of enhanced brightness (intensity), compared to the background, and fits a Gaussian to measure the central location. It then builds up `threads' by connecting local centres in the time-distance diagram. Since the Fe~\textsc{xiii}~1074.7~nm emission is dominated by the inter-thread plasma in the AR setup, rather than the overdense waveguides, NUWT will track the transverse motions of the bright inter-thread regions rather than the threads themselves. In the QS setup, where the overdense threads produce stronger emission than the background, NUWT instead tracks the motions of the waveguides directly. Consequently, the two setups yield fundamentally different physical interpretations of the tracked wave signatures. The bottom panel of Figure \ref{fig:NUWT_tds} shows the Doppler velocity with the same threads over plotted for reference. To extract information on the small-scale torsional waves, let us concentrate on single threads from each simulation, outlined by the cyan curves within the dashed lines in Figure \ref{fig:NUWT_tds}. 

\subsection{Residual Doppler Velocity}

\begin{figure}   
\centering
\begin{subfigure}{0.48\textwidth}
    \includegraphics[width=0.95\textwidth]{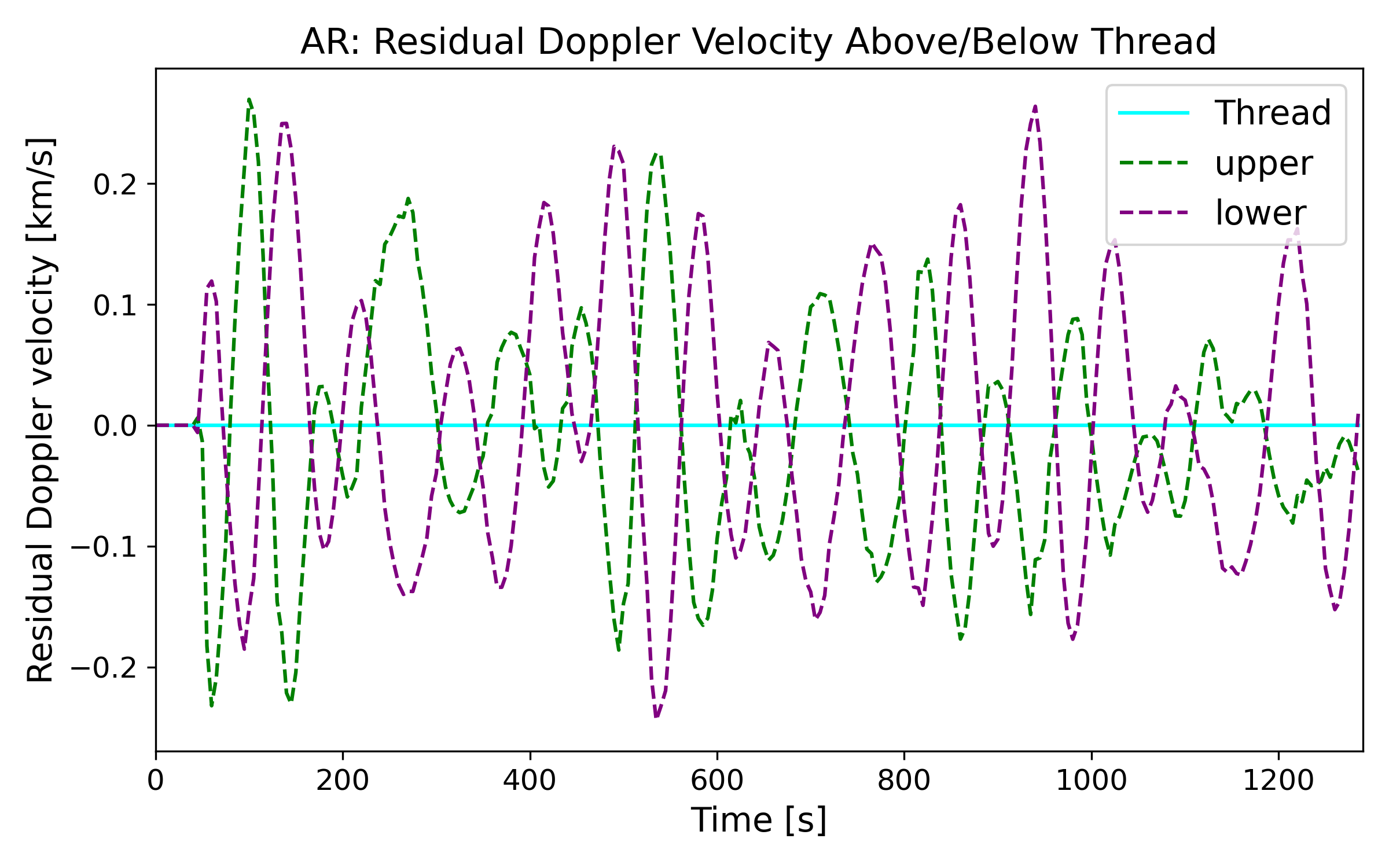}
    \caption{ }    
    \label{fig:residual_doppler_AR}
\end{subfigure}
\begin{subfigure}{0.48\textwidth}
    \includegraphics[width=0.95\textwidth]{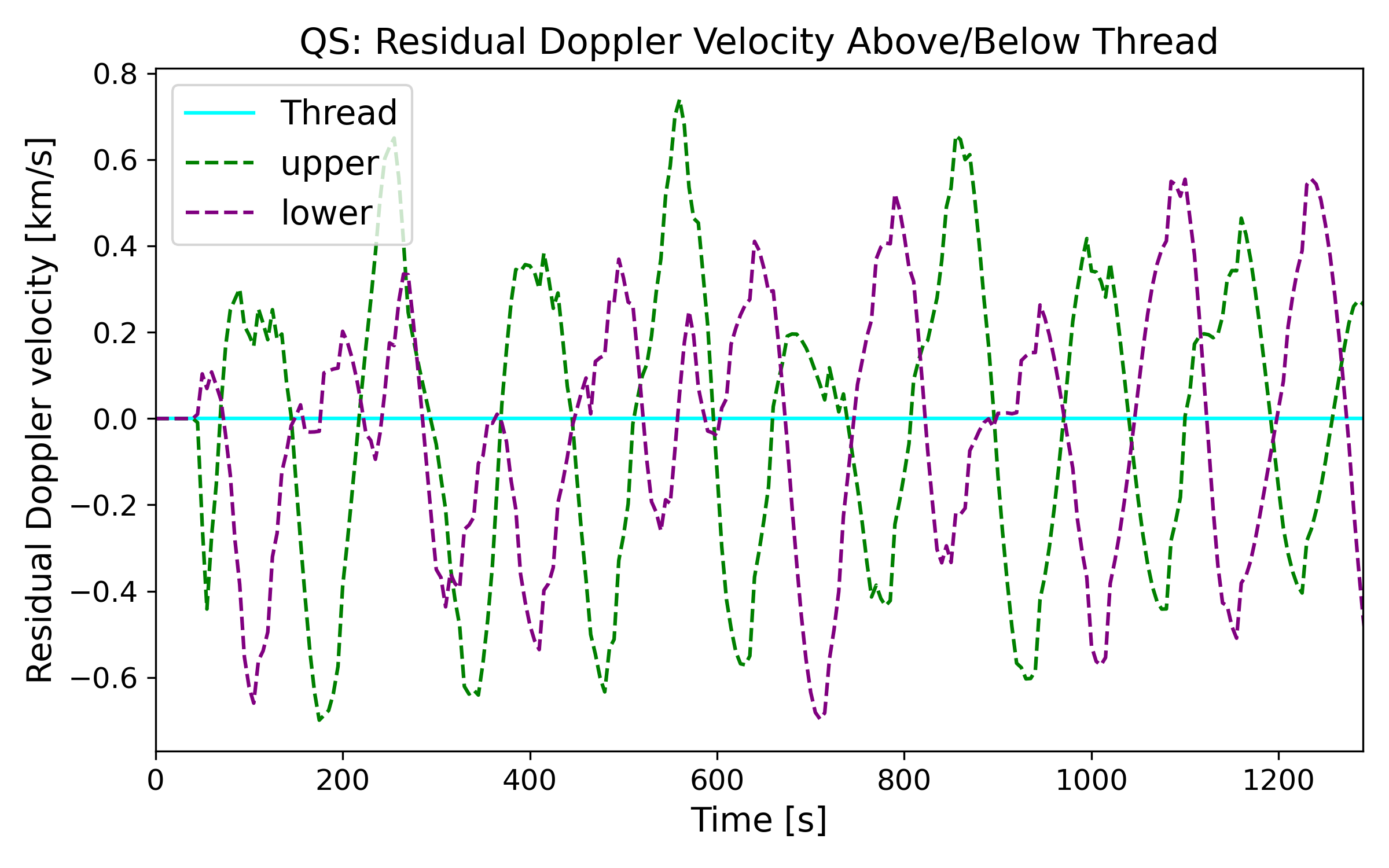}
    \caption{}
    \label{fig:residual_doppler_QS}
\end{subfigure}
\caption{Residual Doppler velocities at the locations of the green and purple curves in Figure \ref{fig:NUWT_zoom_tds} for (a) the AR simulation and (b) the QS simulation. The dashed purple curves denote the residual Doppler velocity below the thread (on one side of the flux tube) whereas the green dashed curves denote the residual Doppler velocity at the same separation above the thread (on the opposite side of the waveguide). The cyan curve denotes the residual Doppler velocity at the thread centre, which is zero by definition.} \label{fig:residual_doppler}
\end{figure}

Figure \ref{fig:NUWT_zoom_tds} displays zoomed versions of the single threads identified for further analysis. The residual Doppler velocity is computed by subtracting the Doppler velocity at the centre of the identified thread from the total Doppler velocity everywhere in the time-distance diagram. As outlined in \citet{Morton2026}, the kink waves produce a bulk motion of the loop threads, and the amplitude of the transverse motion remains roughly constant across each thread. Therefore, in principle, subtracting this bulk velocity reveals the small-scale motions resulting from the torsional waves. The bottom panels of Figure \ref{fig:NUWT_zoom_tds} shows periodic red-blue asymmetries on either side of the centre of the thread (cyan curve) for both simulations. The corresponding time-series for the residual Doppler velocity at $\pm150$~km from the feature centre are shown for both setups in Figure \ref{fig:residual_doppler}.

\begin{figure}   
\centering
\begin{subfigure}{0.48\textwidth}
    \includegraphics[width=0.95\textwidth]{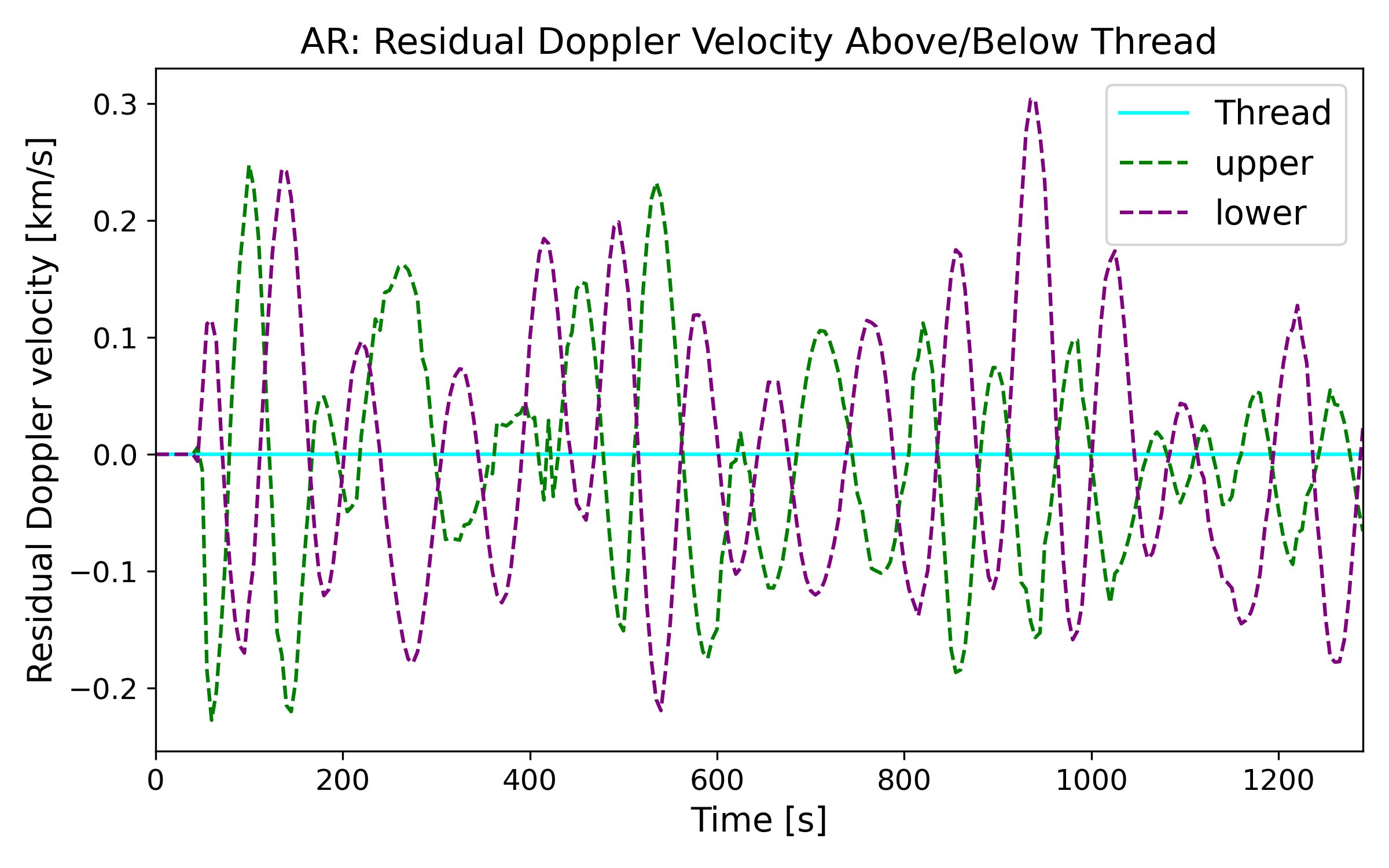}
    \caption{ }    
    \label{fig:residual_doppler_AR_kink_only}
\end{subfigure}
\begin{subfigure}{0.48\textwidth}
    \includegraphics[width=0.95\textwidth]{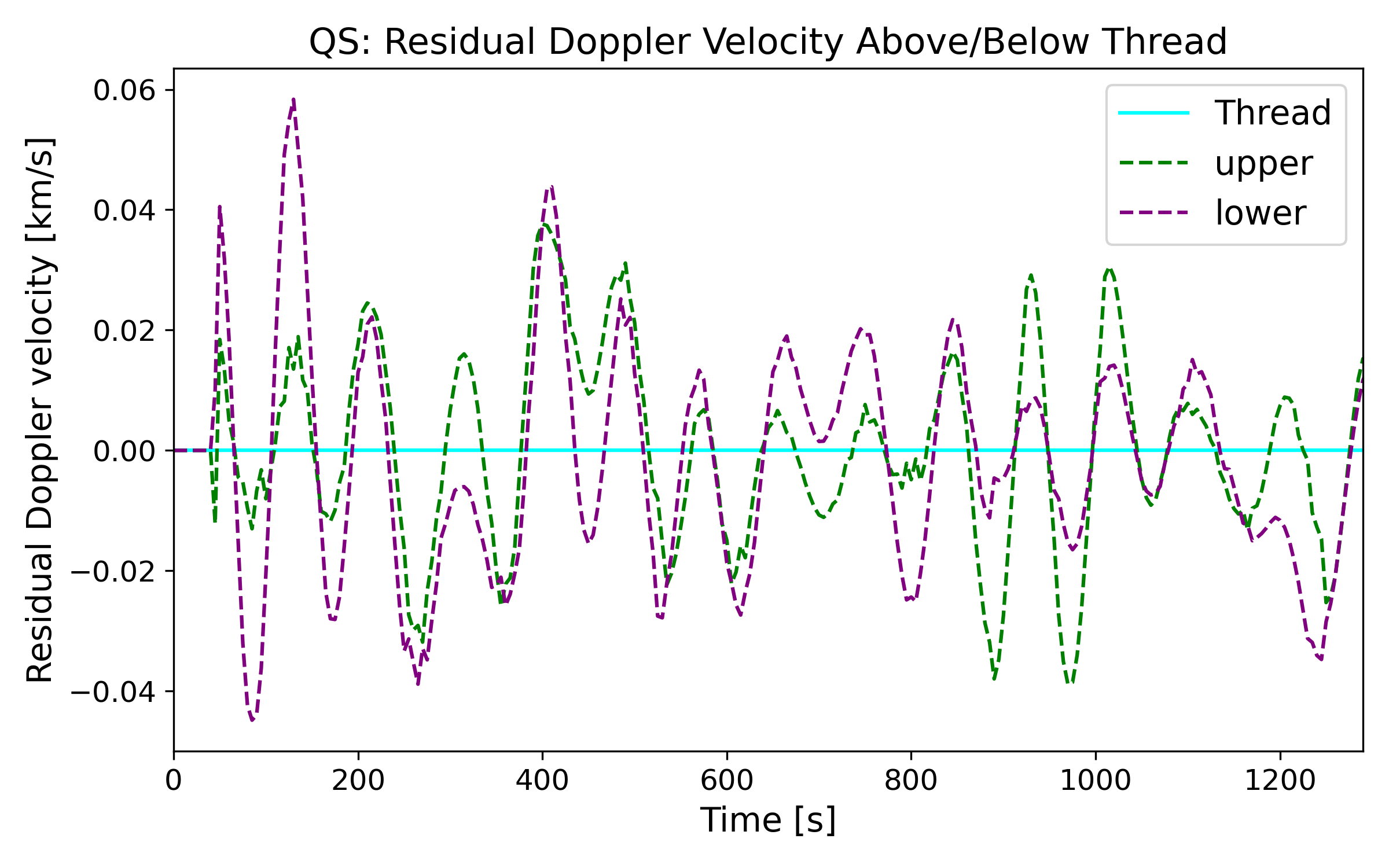}
    \caption{}
    \label{fig:residual_doppler_QS_kink_only}
\end{subfigure}
\caption{Same as Figure \ref{fig:residual_doppler} but for simulations only considering a transverse wave driver. Panel (a) is for the AR simulation whereas panel (b) displays the QS simulation.} \label{fig:kink_only_residual_doppler} 
\end{figure}

Of particular note, anti-symmetric Doppler velocities are visible across the traced intensity feature in the AR simulation, even though the Fe~\textsc{xiii} 1074.7~nm emission from the overdense flux tubes is extremely weak (see Figure \ref{fig:initial_setup}). To confirm the origin of this signal, we run a separate simulation with the same initial setups, however the torsional wave driver is omitted - leaving only the transverse Alfv\'{e}nic motions. The result is displayed in Figure \ref{fig:kink_only_residual_doppler}.

\subsubsection{Quiet Sun}

Figure \ref{fig:residual_doppler_QS_kink_only} displays the residual Doppler analysis for the QS simulation with only a transverse wave driver. The out of phase motions on either side of the identified thread are not present, with the time-series showing in-phase behaviour. This suggests that either (i) there is residual kink wave velocity as the maximum amplitude may not have been at the centre of the thread or (ii) provides evidence for m=1 torsional Alfv\'{e}n waves as a result of resonant absorption in the non-uniform thread boundaries. Nonetheless, the amplitude of the residual Doppler in the QS simulation is extremely small (on the order of $40$~m/s) which indicates that the red/blue asymmetry in Figure \ref{fig:residual_doppler_QS} arises from the driven torsional waves. 

For the residual Doppler velocity in the quiet Sun simulations with the torsional Alfv\'{e}n wave driver, the reduced wave amplitude in the Doppler velocities is consistent with the observational findings of \citet{Morton2026}. It can be understood by considering the cross-sectional velocity structure of each mode. Kink modes perturb the bulk plasma uniformly across the flux tube cross-section, whereas torsional modes have negligible motion at the centre,  with amplitude increasing towards the outer regions of the waveguide, distributed across individual magnetic surfaces. Consequently, when integrated along the line of sight, a greater volume of plasma contributes to the Doppler signal of the kink mode than the torsional mode, resulting in the comparatively weaker torsional wave signature. The decrease in torsional wave amplitude is significant, for example, the driven torsional waves had amplitudes on the order of $16-24$~km~s$^{-1}$, whereas, the `observed' torsional wave amplitudes, determined through the residual Doppler velocity, have dropped to $\approx 0.2-0.6$~km~s$^{-1}$, nearly two orders of magnitude smaller than the true values. This is especially striking when we consider that the Doppler velocity is calculated by integrating across only $6$~Mm in our simulation, whereas the path length through the corona is likely 100's~Mm, suggesting that the amplitude reduction in real observations may be even more severe. We note that, \citet{Pant2019} demonstrate that the amplitude reduction approaches a lower limit for the case of kink modes, with velocities plateauing beyond path lengths of $\sim 100$~Mm, suggesting that the observed amplitudes do not decrease indefinitely with increasing LOS depth. However, properly quantifying the LOS integration effect specifically for torsional modes should be addressed in the future.

Figure \ref{fig:kink_vels_QS} shows the velocity and vertical vorticity ($\omega_z = \frac{\delta v_y}{\delta x} - \frac{\delta v_x}{\delta y}$) field for the simulation driven only with transverse waves. It is evident that the QS setup retains the coherent bulk transverse motions at $z=43.2$~Mm. Although a non-uniformity in the Alfv\'{e}n speed is present, the lower density contrast produces only weak phase mixing, insufficient to generate significant out-of-phase motions on the timescales simulated. The vorticity panel of Figure \ref{fig:kink_vels_QS} reflects this clearly as $\omega_z$ is substantially weaker in magnitude for the QS setup than in the AR case, and the horizontal vorticity arrows show a more organised pattern that is correlated with the locations of individual waveguides. The emission is also distributed more uniformly across the domain rather than being concentrated at waveguide boundaries, a direct consequence of the weaker density structuring. As a result, the bulk Doppler velocity at the centre of the waveguide faithfully represents the driven transverse waves, and subtracting it effectively isolates any superposed torsional perturbation. The QS setup therefore provides a more favourable environment for disentangling torsional Alfv\'{e}n waves from global transverse motions using residual Doppler diagnostics, precisely because the low density contrast suppresses the phase mixing and velocity shear that would otherwise contaminate the torsional-like signal.

\subsubsection{Active Region}\label{sec:AR}

Let us now turn our attention to the AR simulation. Figure \ref{fig:residual_doppler_AR_kink_only} displays the residual Doppler motions on either side of an isolated flux tube for the AR setup with only a transverse wave driver. Surprisingly, we still see the same signatures of anti-symmetric Doppler motions on opposite edges of the flux tube. It is clear that these motions do not arise from the driven torsional waves, as the torsional wave driver is not present in this case. This raises a few interesting questions, namely: (i) what is the origin of the $m=0$ torsional-like signatures in the AR simulation when only a kink wave driver is included?; (ii) how do we conclusively identify spectroscopic signatures of pure $m=0$ torsional Alfv\'{e}n waves? 

\begin{figure*}
\centering
\begin{subfigure}{0.98\textwidth}
    \includegraphics[width=\textwidth]{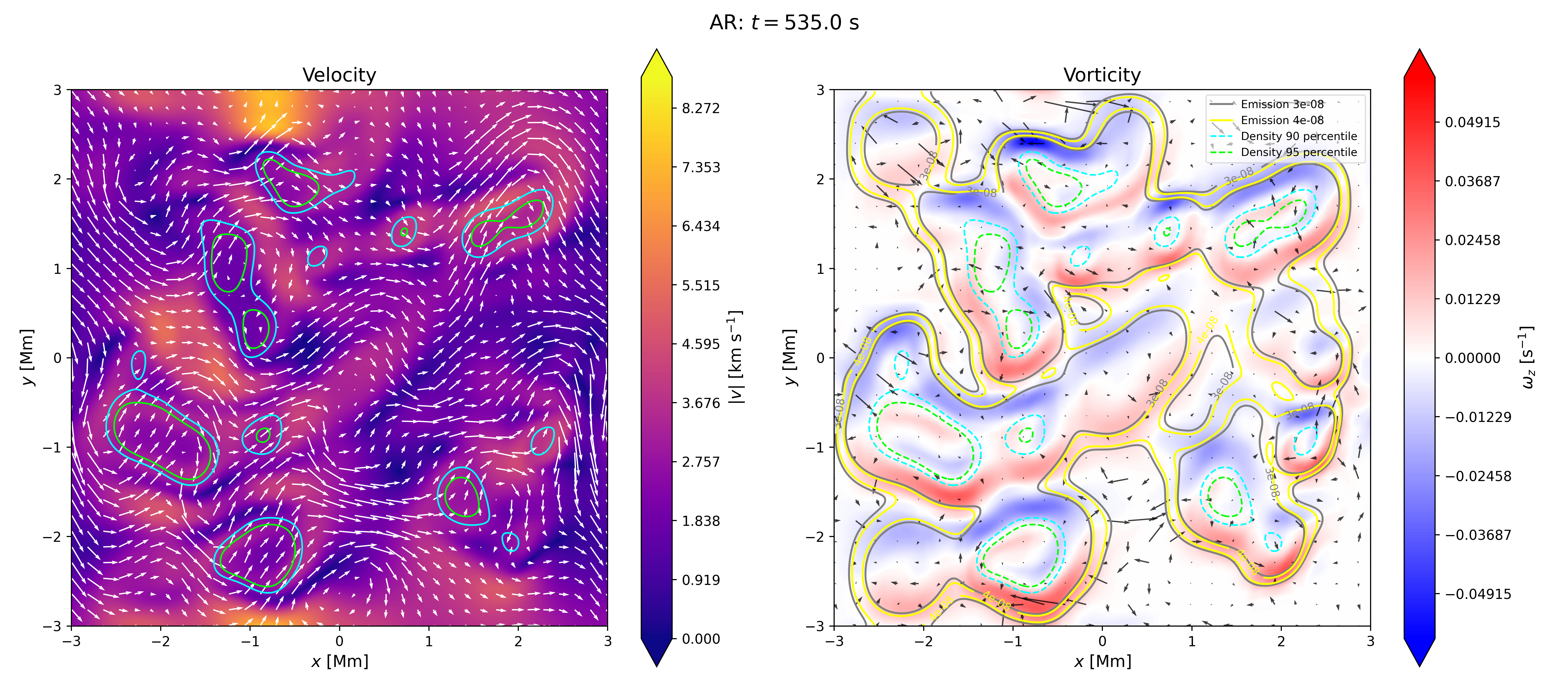}
    \caption{Kink Only AR simulation}
    \label{fig:kink_vels_AR}
\end{subfigure}
\begin{subfigure}{0.98\textwidth}
    \includegraphics[width=\textwidth]{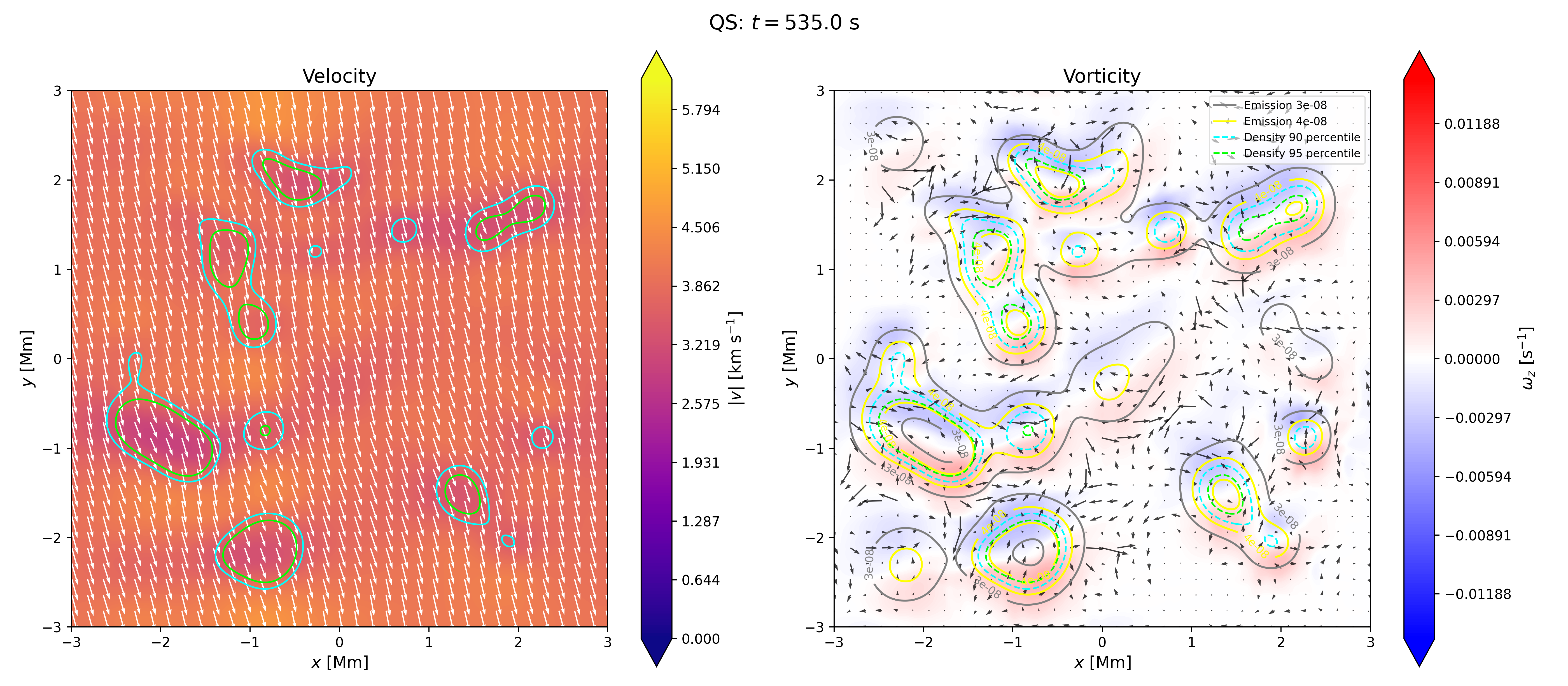}
    \caption{Kink Only QS simulation}
    \label{fig:kink_vels_QS}
\end{subfigure}
\caption{Horizontal slices at $z=43.2$~Mm of the transverse velocity amplitude (left panels) and vertical vorticity $\omega_z$ (right panels) at $t=535$~s, for the AR setup (a) and QS setup (b). In the left panels, white arrows show the horizontal velocity vector field and the cyan/lime dashed contours outline the locations at $90\%$ and $95\%$ of the maximum density, respectively. In the right panels, black arrows show the approximate horizontal vorticity ($\omega_x, \omega_y$), and the cyan/lime dashed contours again mark the density thresholds. The grey and yellow solid contours indicate regions of Fe~\textsc{xiii} 1074.7~nm emission at $3\times10^{-8}$ and $4\times10^{-8}$~erg~sr$^{-1}$~cm$^{-3}$~s$^{-1}$, respectively. Note that the vorticity vectors are scaled differently for both setups for visualisation purposes.}
\label{fig:vels_kink_only}
\end{figure*}

It is worth identifying why the red/blue asymmetric signal appears in the AR setup with no torsional wave driver. Figure \ref{fig:kink_vels_AR} shows that the plasma in the AR-like setup exhibits a strongly phase-mixed, shear-dominated character at this height. This can be understood through the process of phase mixing. As shown in Figure \ref{fig:valf_comparison}, there exists a strong gradient in the Alfv\'{e}n speed across the magnetic field in the AR setup due to the high density contrast between the overdense waveguides and the surrounding plasma. As a result, the driven transverse (kink) waves, whose phase speed is determined by the local kink speed, propagate at different speeds along the magnetic field. By the time they reach $z=43.2$~Mm, the threads are oscillating out-of-phase with one another. These out-of-phase transverse motions interact across waveguide boundaries, generating velocity shear and further contributing to the turbulent-like state of the plasma.

The vorticity panel of Figure \ref{fig:kink_vels_AR} illustrates this picture directly. The vertical vorticity displays large-scale, spatially incoherent patches of positive and negative rotation that fill the volume in-between the waveguides. Crucially, these patches bear no organised relationship to the individual waveguide boundaries identified by the density contours, in contrast to the $\omega_z$ pattern that would be expected around each waveguide in the case of a genuine torsional Alfv\'{e}n wave. Whilst the red/blue asymmetries displayed by the vertical vorticity do not immediately represent what would be observed by the LOS integrated Doppler velocity, they do indicate rotational and/or shear motions which may be confused with genuine torsional waves. Furthermore, the horizontal vorticity arrows ($\omega_x, \omega_y$) point in spatially random directions for the AR simulation, again inconsistent with the azimuthally organised horizontal vorticity signature of a torsional wave. Notably, the emission contours concentrate beyond the waveguide boundaries, precisely where the velocity shear is strongest, meaning that any line-of-sight integration of emission-weighted Doppler velocity will be dominated by these shear flows rather than by coherent wave motions.

This turbulent-like vorticity structure is crucial for interpreting the residual Doppler analysis displayed in Figures \ref{fig:NUWT_td_zoom_AR} and \ref{fig:residual_doppler_AR}. The residual Doppler technique relies on isolating torsional wave signatures by subtracting the bulk transverse wave velocity at the centre of the identified waveguide. However, in the AR setup this approach breaks down for two reasons. First, the centre of the identified thread resides in a location between waveguides rather than within one, so the reference velocity does not represent a clean bulk transverse motion. Second, the shear and rotational flows generated by the phase-mixed plasma create $m = 0$-like signatures in the residual Doppler velocity even in the complete absence of a torsional wave driver, as demonstrated in Figure \ref{fig:residual_doppler_AR_kink_only}. In other words, the shear flows mimic the observational signature of torsional Alfv\'{e}n waves.

\section{Discussion and Conclusions}\label{sec:conclusions}

In this study we have performed 3D numerical simulations of transverse and torsional waves on inhomogeneous density enhancements representing coronal waveguides. We conducted forward modelling of the Fe~\textsc{xiii} 1074.7~nm line to investigate the spectroscopic signatures of the driven waves observed by DKIST. We created two numerical setups, by varying the strength of the density inhomogeneities, designed to represent plasma under quiet Sun and active region conditions. The transverse waves are global across the whole domain, whereas the torsional wave drivers are localised specifically to the density enhancements which act as MHD waveguides.

One significant aspect this study reveals is that the interpretation of features in coronal line emission diagnostics depends on the underlying inhomogeneity in the different regions of the atmosphere. Observations indicate that the density inhomogeneity in active regions is at least twice that in the quiet Sun, which is represented in our modelling. While the two different plasma conditions lead to diagnostics with similar qualities, the LOS integrated emission does not correspond to the same plasma features. For example, the forward modelled line amplitude indicates bright and dark features in the time-distance maps, but these only map to waveguides in the quiet Sun setup. Assuming a constant field strength across the local region, the presence of the density inhomogeneity leads to differences in temperatures of the plasma associated with a waveguide and the ambient plasma. Hence the emission from active regions likely forms in compact regions restricted to plasma with temperatures close to the peak formation temperature of the ion (in this case Fe~\textsc{xiii}). In the model we use here, this means the strongest regions of emission are not associated with the waveguides, but in volumes around and between the waveguides. The waveguides themselves have only a small contribution to the emergent intensity, approximately five orders of magnitude smaller than the strongest emission. In the quiet Sun case, the density inhomogeneity is small and the emission from the waveguides is comparable to the plasma with the largest emission (Figure~\ref{fig:initial_setup}). Hence the emergent intensity is weighted more towards the waveguides, with bright features seen in the line amplitude data corresponding to waveguides.

\medskip

This feature of the emergent intensity influences any subsequent inferences from an analysis of wave motions. Following, what is standard practice in the analysis of observational data, we created time–distance maps of the forward-modelled intensity output and analysed the wave behaviour along regions of enhanced emission. In both simulations, the amplitudes and periods of the kink modes are well recovered from the displacements associated with the motion of the bright threads in the POS. This is because the wave field associated with the kink motion is coherent over the region and all the plasma is perturbed in the same way. The kink motions are known to be coherent over patches of around $7-8$~Mm in the corona \citep{Sharma2023}, so one should expect the observed POS motions, and also the Doppler velocities, correspond to the wave field associated with the kink motion. The interpretation of signatures of  torsional motions is more nuanced.

\medskip

To examine the presence of torsional motions in the emergent intensity, we followed \cite{Morton2026} and subtracted a signal assumed to be related to the bulk transverse kink motion from the Doppler velocity, isolating the residual Doppler signals. For the QS simulation setup, we recovered the signature of the $m=0$ torsional Alfv\'{e}n waves propagating along the density enhancements, with antisymmetric Doppler velocities on opposite sides of the identified structures. The `observed' amplitude of the torsional waves had reduced by two orders of magnitude compared to the driven waves, as a result of line-of-sight integration effects through the plasma. Moreover, in the absence of the torsional wave driver, the Doppler motions on opposite sides of the waveguides exhibit an in-phase relationship with small amplitudes ($<50$~m~s$^{-1}$), indicative of residual kink motions or resonant absorption.

In contrast, red-blue Doppler asymmetries, commonly interpreted as signatures of $m=0$ torsional waves, are also found in the AR setup. However, they are present both with and without the torsional Alfv\'{e}n wave driver. This suggests that the origin of the torsional motions in the AR case is not wholly due to the driven torsional waves. This phenomenon occurs because the plasma dominating the emergent intensity is not associated with the waveguide. The driven torsional modes are localised to the  waveguides \citep[in line with observational indications from][]{Morton2026}, hence only have a very weak effect on the emergent intensity. Examination of the velocity and vorticity fields in the inter-waveguide regions shows evidence of turbulent-like plasma motions, producing shear and rotational flows in the locations of strongest emission, producing red/blue Doppler asymmetries on opposite sides of identified flux tubes which are easily interpreted as signatures of torsional waves. 

Hence, this case study highlights that care should be taken when analysing residual Doppler velocities at opposite edges of coronal structures when searching for torsional Alfv\'{e}n waves. Although antisymmetric Doppler motions are a necessary signature of $m=0$ torsional modes, they are not by themselves sufficient in strongly inhomogeneous plasma. However, when applied to plasma where the density inhomogeneity is small, residual Doppler analysis remains a powerful and reliable method for identifying torsional Alfv\'{e}n waves. This supports the results presented in \cite{Morton2026}, which focused on fine-scale structure in the quiet Sun which appears to be weakly inhomogeneous \citep[e.g.,][]{Morton2021_QS_damping}.

\medskip
Our results naturally raise the question of how one may conclusively identify spectroscopic signatures of genuine $m=0$ torsional Alfv\'{e}n waves in the presence of phase-mixed shear flows. Several potential discriminating diagnostics can be explored. First, a Fourier analysis of the residual Doppler time series could reveal a spectral peak at the torsional wave frequency if it is distinct from the kink wave spectrum. In the present work the torsional driver has a fixed period of $P=150$~s which is separable from the broadband kink spectrum, but in reality torsional wave driving is expected to be broadband, mainly driven by granulation and convective motions just like the kink waves, making spectral separation non-trivial. Second, the synthesised line width may carry information about the wave environment. Even at the native simulation resolution, torsional waves contribute to line broadening through the spatial gradient in the velocity field across the flux tube, producing a spread in Doppler shifts that broadens the line profile. A torsional wave is expected to produce in-phase broadening at opposite edges of a flux tube with minimal broadening at the centre, providing a potential discriminating diagnostic. We have carried out a preliminary analysis of the synthesised line width in all four simulations, however, the physical interpretation of the line width signatures in the context of torsional wave identification remains unclear and will be the subject of future work. Third, height-resolved spectroscopy could, in principle, reveal the propagating nature of torsional waves through a systematic phase lag in the residual Doppler signal with height. Genuine torsional Alfvén waves propagate at the Alfv\'{e}n speed, whereas the phase-mixed red-blue asymmetries in the AR case are driven by the kink waves and would be expected to propagate at the kink speed. Therefore, comparing the propagation speed of the red-blue asymmetry to the independently measured kink wave propagation speed could discriminate between the two cases. However, we note that in the quiet Sun corona the density contrast between waveguides and the surrounding plasma is small ($\rho_i/\rho_e \approx 1$), meaning the kink speed and Alfv\'{e}n speed are nearly identical and this diagnostic would be difficult to apply in practice in regions with weak density structuring.

\medskip

The modelling here comes with some caveats. In the present work, the transverse kink waves are driven with a broadband velocity spectrum, with amplitudes and frequencies drawn from an observed distribution, whereas the torsional wave drivers are prescribed with a fixed period. Furthermore, the torsional wave drivers are present throughout the duration of the simulation, whereas \citet{Morton2026} reported that the signatures of torsional waves were coherent for only a period or two. In reality, the solar corona is subject to a more complex, non-uniform driving at the photospheric boundary. Observations of magnetic bright points in the intergranular lanes reveal that magnetic flux tube footpoints undergo a combination of transverse buffeting and rotational motions, which are expected to generate a mixture of kink and torsional Alfv\'{e}nic waves simultaneously \citep{Jess2009, Utz2010}. Furthermore, whilst our driving is applied uniformly across the lower boundary on coherent length scales, photospheric motions are inherently localised and non-uniform. The extent to which a broadband torsional driver, or a more spatially non-uniform boundary prescription, would modify the resulting wave signatures, including the Doppler velocity amplitudes and power spectra, remains an open question. In particular, driving the torsional waves with a power-law frequency distribution may alter the relative energy partition between the two wave modes, with implications for their observational signatures. The results presented here are based on a single random realisation of the density field for each setup. We argue that the key conclusions are governed by the bulk plasma properties rather than the specific spatial configuration of the waveguides. In particular, for emission to be associated with the overdense waveguides, the plasma density and temperature must be favourable to the formation of the Fe~\textsc{xiii} line, which is determined by the thermodynamic conditions rather than the spatial arrangement of the enhancements. Nonetheless, repeating the analysis for multiple random realisations to confirm the generality of the results remains a valuable avenue for future work.

In addition, realistic coronal observations sample multiple unresolved structures along the line of sight (with path lengths at least 300~Mm), each potentially hosting incoherent wave motions. Future forward-modelling efforts should therefore consider the influence of the superposition of multiple structures over longer path lengths, quantifying how incoherent torsional motions manifest in Doppler and line-width diagnostics. This will require determining the appropriate emission weighting of structures along the LOS, particularly given that photoexcitation of the Fe~\textsc{xiii} 1074.7~nm line depends on the height of the emitting plasma above the limb. Such modelling will be crucial for interpreting observed non-thermal broadening and for disentangling genuine wave signatures from LOS superposition effects.

Finally, the present work potentially motivates future multi-spectral studies of torsional Alfv\'{e}n waves. Since the observable signatures in our simulations are sensitive to where emission is strongest, governed by the density and temperature structure of the waveguides, spectral lines formed at different temperatures and densities may show unique signatures. In particular, cooler lines such as Fe~\textsc{xi}~$789$~nm ($\log T \sim 6.1$) may preferentially sample the interior of the waveguides rather than the  plasma in-between in our AR setup, potentially recovering greater sensitivity to true torsional motions distinct from the phase-mixed vortical flows. Such multi-spectral observations are currently limited by instrument capabilities, however, the upcoming DL-NIRSP instrument will enable simultaneous observations of multiple coronal lines, and Solar-C/EUVST will provide multi-thermal EUV spectroscopy, offering promising avenues for disentangling the competing wave signatures identified in this work.

\begin{acknowledgments}
We thank the anonymous referee or their insightful comments and suggestions. S.J.S and R.J.M are grateful for support from the UKRI Future Leader Fellowship Grant (RiPSAW MR/Z000289/1). This work used the Oswald High Performance Computing facility operated by Northumbria University (UK), and the DiRAC Memory Intensive Service (Cosma7) at Durham University, managed by Durham University on behalf of the STFC DiRAC HPC Facility (www.dirac.ac.uk). The DiRAC component of Cosma7 at was funded by UKRI and STFC capital funding, as well as STFC operations grants. DiRAC is part of the UKRI Digital Research Infrastructure. The data behind figures and the simulation setup are available in Figshare at DOI: 10.25398/rd.northumbria.32907107. Additional data supporting the findings of this study are available from the corresponding author upon reasonable request. 
\end{acknowledgments}

\software{Data analysis has been undertaken with the help of NumPy \citep{Numpy}, 
matplotlib \citep{Matplotlib}, and SciPy \citep{Scipy}.}

\bibliography{ref}{}
\bibliographystyle{aasjournal}

\end{document}